\documentclass[aps,prl,nofootinbib,longbibliography,superscriptaddress,twocolumn]{revtex4-2}
\usepackage{amsmath}
\usepackage{amssymb}
\usepackage{graphicx}
\usepackage{dcolumn}
\usepackage{bm}
\usepackage[colorlinks=true,linkcolor=blue,anchorcolor=blue, citecolor=blue,urlcolor=blue]{hyperref}
\usepackage[mathlines]{lineno}
\usepackage{ulem}
\usepackage{siunitx}
\usepackage{gensymb}

\begin{document}
\title{Odd-Parity Magnons}
\author{Pu Zhang}
\author{Sun-Bo Xie}
\author{Junxi Yu}
\author{Yichen Liu}
\author{Cheng-Cheng Liu}
\email{ccliu@bit.edu.cn}
\affiliation{Centre for Quantum Physics, Key Laboratory of Advanced Optoelectronic Quantum Architecture and Measurement (MOE), School of Physics, Beijing Institute of Technology, Beijing 100081, China}

\begin{abstract}

Magnons, as charge-neutral spin excitations, can transport spin information without Joule heating and therefore offer a promising platform for low-power spintronics. However, in collinear magnets, the effective time-reversal symmetry forbids odd-parity magnon band splitting. Here we propose odd-parity magnons and establish a general mechanism for realizing them in collinear antiferromagnets. We provide a complete spin-point-group classification of odd-parity magnon splitting in two-dimensional collinear antiferromagnets by identifying the leading splitting types and their symmetry-allowed basis functions. This classification serves as a practical guide for searching for odd-parity magnons. We show that breaking effective time-reversal symmetry, for example by circularly polarized light or loop currents, can induce highly tunable $p$- and $f$-wave magnon splitting. In bilayer systems, the dynamical modulation can drive a topological magnon phase transition, accompanied by chiral edge modes and an abrupt jump in the magnon thermal Hall conductivity. Material-specific first-principles calculations further demonstrate the feasibility of this mechanism in real van der Waals antiferromagnets.  Our study identifies the odd-parity magnons as a new class of spin excitations and provides a theoretical foundation for odd-parity magnons and ultrafast optically controlled topological magnonic devices.
\end{abstract}

\maketitle

\textit{Introduction.}---Spin splitting in momentum space has emerged as a central concept for understanding and engineering novel magnetic states~\cite{yamada2025metallic,hellenes2023p,lin2025odd,ezawa2025third,brekke2024minimal,yu2025odd,song2025electrical,zhuangOddParityAltermagnetismOriginated2025,panOrbitalAltermagnetism2025,huangLightInducedOddParityMagnetism2026,zhuFloquetOddParityCollinear2026,Li2026Floquet,liuLightinducedOddparityAltermagnets2026,zhu2025design,wu2025magnon,jin2026interaction,cui2023efficient,chen2025unconventional,karaki2025high,yu2026sliding}. A representative example is provided by altermagnets, which combine symmetry-protected reciprocal-space spin splitting with a fully compensated collinear magnetic order in real space~\cite{smejkalEmergingResearchLandscape2022,mazinEditorialAltermagnetismNew2022}. Beyond fermionic quasiparticles, magnetic systems also host bosonic spin excitations, namely magnons. Because magnons are charge-neutral, they can transport spin angular momentum without Ohmic losses, making them attractive for low-power spintronic and magnonic devices~\cite{chumak2015magnon,khitunMagnonicLogicCircuits2010,baoDiscoveryCoexistingDirac2018,zhang2013topological,shindou2013topological,mook2014edge,li2018topological,lu2021topological,neumann2022thermal,hu2022tunable,bai2026ferroelectrics}.  In collinear magnets, however, the intrinsic spin-group symmetry, or equivalently the effective time-reversal symmetry, constrains both electronic and magnonic band structures to exhibit only even-parity spin splitting, such as $d$-, $g$-, or $i$-wave forms. Realizing odd-parity magnon spin splitting, such as $p$- or $f$-wave splitting, in collinear magnets therefore remains a fundamental challenge.

\begin{figure}[htbp]
	\centering 
	\includegraphics[width=0.8\linewidth]{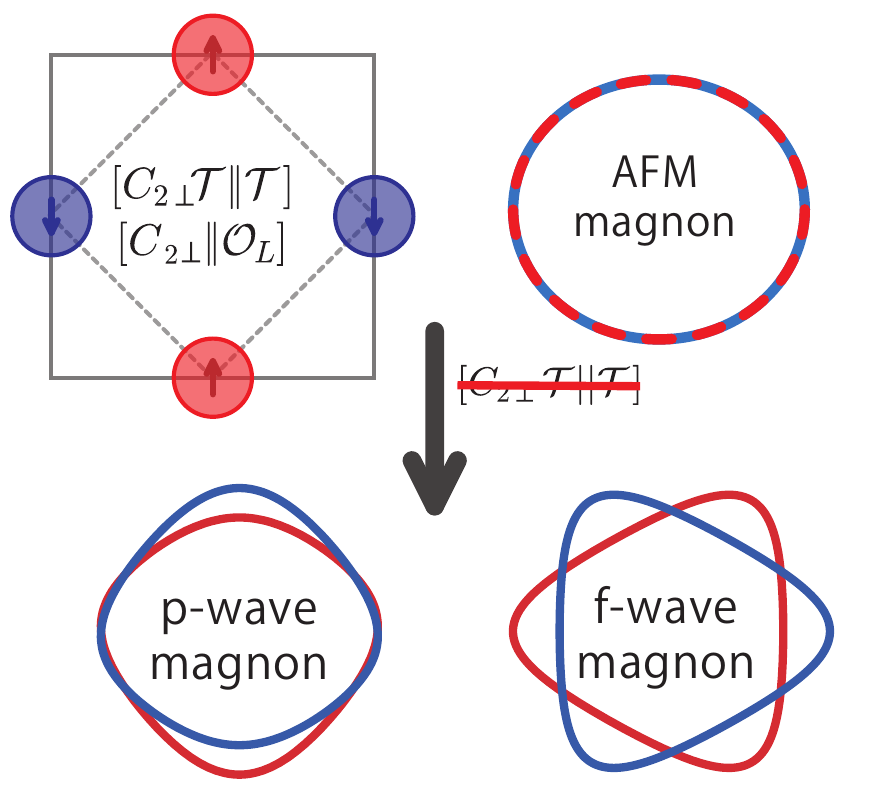}
	\caption{Schematic of the physical mechanism for odd-parity magnons in collinear antiferromagnets. (Top) A collinear antiferromagnetic lattice has the joint symmetries $[C_{2\perp}\mathcal{T}||\mathcal{T}]$ and $[C_{2\perp}||O_L]$ with $O_L=P$, $m$, $C_{2z}$, $S^{3}_{3z}$, or $C^{3}_{6z}$. The red and blue spheres denote the two sublattices with spin-down and spin-up polarizations, and the magnon branches are spin-degenerate. Breaking the effective time-reversal symmetry $[C_{2\perp}\mathcal{T} || \mathcal{T}]$, for example by circularly polarized light or loop currents, while preserving $[C_{2\perp}|| O_L]$, induces odd-parity magnon spin splitting. (Bottom) Odd-parity magnon states dictated by different lattice symmetries, illustrating the $p$-wave  (left) and $f$-wave (right) spin-splitting distribution on a constant-energy contour.}
	\label{pic1}
\end{figure}

In this work, we propose odd-parity magnons and develop a microscopic theory for their realization in collinear antiferromagnets. As schematically illustrated in Fig.~\ref{pic1}, odd-parity magnon splitting can be generated by breaking the effective time-reversal symmetry $[C_{2\perp}\mathcal{T} || \mathcal{T}]$ while preserving the symmetry $[C_{2\perp} || O_L]$, where $O_L$ is a momentum-reversing lattice operation. This symmetry criterion can be implemented, for example, by circularly polarized light (CPL)~\cite{oka2019floquet,bao2022light} or loop currents~\cite{bourges2021loop}, which break $\mathcal{T}$ without destroying the relevant lattice symmetry. We further provide a complete spin-point-group classification of odd-parity magnons in two-dimensional collinear antiferromagnets, including the leading splitting types and their symmetry-allowed basis functions, as summarized in Table~\ref{table:odd_parity_2D}. In monolayer systems, we show that CPL or elliptically polarized light can induce highly tunable $p$- and $f$-wave magnon splitting. In bilayer systems, the same dynamical modulation can drive a topological magnon phase transition, producing chiral edge modes and an abrupt jump in the magnon thermal Hall conductivity. First-principles-based calculations for monolayer MnPS$_3$, bilayer FeBr$_3$, slid bilayer CrI$_3$, and CrVI$_6$ further demonstrate the feasibility of realizing odd-parity magnons in realistic van der Waals antiferromagnets.

\textit{Symmetry Analysis.}---In the nonrelativistic limit, the spin and crystalline degrees of freedom are decoupled. A symmetry operation of a magnetic system can therefore be written as a direct product of a spin-space operation and a real-space operation, denoted by spin-group operation $[R_S || R_L]$. In collinear magnets, the effective time-reversal symmetry $[C_{2\perp}\mathcal{T} || \mathcal{T}]$ enforces an even-parity dispersion $\omega(s,\mathbf{k})=\omega(s,-\mathbf{k})$, where $s=\uparrow,\downarrow$ labels the magnon spin branchs. In contrast, noncoplanar magnets, which lack this symmetry, can host odd-parity magnon dispersions~\cite{neumann2026odd}.  If the system further possesses a spin-group symmetry $[C_{2\perp} || O_L]$, where $O_L$ is a lattice operation that reverses momentum, such as inversion $P$ or a twofold rotation $C_{2z}$ in two dimensions, the magnon spectrum also satisfies $\omega(s, \mathbf{k}) = \omega(-s, -\mathbf{k})$. The coexistence of $[C_{2\perp}\mathcal{T} || \mathcal{T}]$ and $[C_{2\perp} || O_L]$ then enforces a global spin degeneracy throughout the Brillouin zone, $\omega(s,\mathbf{k})=\omega(-s,\mathbf{k})$, thereby forbidding spin splitting.

This observation leads to a general symmetry criterion for realizing the odd-parity magnons: breaking $[C_{2\perp}\mathcal{T} || \mathcal{T}]$ while strictly preserving $[C_{2\perp} ||O_L]$. Physically, $\mathcal{T}$ can be effectively broken by means such as CPL or loop currents. Concurrently, these spatially uniform means do not destroy $O_L$, thus retaining the odd-parity operation $[C_{2\perp} ||O_L]$.
Under these conditions, the spin degeneracy is lifted, but the remaining spin-point-group symmetries continue to constrain the momentum dependence of the magnon spin splitting. These constraints can be formulated directly in terms of the magnon spin-splitting function $\Delta\omega(\bm{k}) \equiv \omega_{\uparrow}(\bm{k})-\omega_{\downarrow}(\bm{k})$. For each point-group operation $g$, the splitting obeys
\begin{equation}
    \Delta\omega(g\bm{k})=\chi(g)\Delta\omega(\bm{k}),
\end{equation}
where $\chi(g)=+1$ for the spin-preserving lift $[E\|g]$ and $\chi(g)=-1$ for the spin-flipping lift $[C_{2\perp}\|g]$ in the collinear spin point group.
Thus $\Delta\omega(\bm{k})$ transforms as a basis function of a one-dimensional representation of the underlying point group.

Following this criterion, we systematically examine the one-dimensional irreducible representations of all point groups and identify the complete set of collinear spin point groups that can support odd-parity splitting. The resulting classification for two-dimensional collinear magnets is summarized in Table~\ref{table:odd_parity_2D}, together with the leading odd-parity splitting type and the corresponding basis functions. In the Floquet case, CPL also induces an effective out-of-plane axial field. This field modifies the symmetry action of in-plane twofold rotations and vertical mirrors, which must be combined with time reversal when they reverse the axial field. Consequently, the allowed basis functions can differ from those in the static case, as summarized in the “Floquet Splitting” column of Table~\ref{table:odd_parity_2D}. We note that the classification in Table~\ref{table:odd_parity_2D} is based only on spin-point-group symmetry and therefore applies not only to bosonic magnon bands, but also to odd-parity electronic spin splittings.

\textit{Floquet Odd-Parity Magnons.}---To verify the aforementioned symmetry criteria and elucidate the microscopic induction mechanism of odd-parity magnons, we construct an effective spin model based on a two-dimensional (2D) collinear antiferromagnet, selecting periodic Floquet optical driving as the specific modulation means. First, consider the unperturbed Heisenberg Hamiltonian
\begin{equation}
	\mathcal{H}_0 = J \sum_{\langle \alpha,\beta \rangle} \mathbf{S}_\alpha \cdot \mathbf{S}_\beta+K \sum_{\alpha} (S_\alpha^z)^2,
\end{equation}
where $\mathbf{S}_\alpha$ and $\mathbf{S}_\beta$ denote the local spin operators at lattice sites $\alpha$ and $\beta$, respectively. $\langle \alpha,\beta \rangle$ represents the summation over all nearest-neighbor (NN) site pairs, and $J>0$ is the exchange constant characterizing the antiferromagnetic coupling strength between nearest-neighbor spins. $K<0$ represents the easy-axis anisotropy, which stabilizes the Néel vector along the $\hat{z}$ direction. This term does not affect the symmetry breaking or the ensuing odd-parity spin splitting. For analytical simplicity, the $K$ term will be omitted in the following discussion. In this ground state, the system belongs to the collinear spin point group $^{\bar{1}}6/m^{\bar{1}}mm$, and is protected by the effective time-reversal symmetry $[C_{2\perp}\mathcal{T} || \mathcal{T}]$ and the joint spin-spatial inversion symmetry $[C_{2\perp} || P]$, maintaining the magnon bands strictly degenerate.

According to our established symmetry mechanism, the key to lifting this degeneracy and inducing odd-parity splitting lies in breaking $\mathcal{T}$. Magnons, which are electrically neutral yet possess a non-zero magnetic moment, can couple to a time-dependent electric field $\mathbf{E}(t)$ via the Aharonov-Casher effect \cite{aharonovTopologicalQuantumEffects1984a}. The electric field is given by $\mathbf{E}(t) \times \hat{z} = E_0\left(\xi\sin\Omega t, -\cos\Omega t, 0\right)$, where $\xi=\pm1$ indicates right- or left-handed CPL. Hereafter, we take right-handed CPL ($\xi=1$) as an example; the left-handed case can be analyzed similarly. During the hopping process, a time-dependent phase $\theta_{\alpha\beta}(t) = \lambda \sin(\Omega t - \phi_{\alpha\beta})$ is acquired, where $\lambda = \frac{g\mu_B E_0 r_{\alpha\beta}}{\hbar c^2}$ acts as a dimensionless coupling strength, and $r_{\alpha\beta}$ is the distance between two sites ~\cite{owerreFloquetTopologicalMagnons2017a,vinasbostromLightinducedTopologicalMagnons2020}. Consequently, the time-dependent Heisenberg Hamiltonian reads
\begin{equation}
	\mathcal{H}(t) = \sum_{\langle\alpha,\beta\rangle} \left[ \frac{J}{2} \left( S_\alpha^- S_\beta^+ e^{i\theta_{\alpha\beta}(t)} + \text{h.c.} \right) + J S_\alpha^z S_\beta^z \right],
\end{equation}
where $S^\pm = S^x \pm i S^y$ are the spin raising and lowering operators. 

To obtain the effective Hamiltonian, we employ the Floquet-Magnus expansion in the high-frequency limit, $\mathcal{H}_{\text{eff}} \approx \mathcal{H}_{\text{eff}}^{(0)} + \mathcal{H}_{\text{eff}}^{(1)}$. The zeroth-order term $\mathcal{H}_{\text{eff}}^{(0)} = \mathcal{H}^0$ solely leads to a renormalization of the exchange coupling strength. The key to breaking the time-reversal symmetry $\mathcal{T}$ lies in the first-order correction, which is given by the commutator of the Fourier components of the Hamiltonian
\begin{equation}
	\mathcal{H}_{\text{eff}}^{(1)} = \sum_{n=1}^{\infty} \frac{1}{n \hbar\Omega}[\mathcal{H}^n, \mathcal{H}^{-n}]+O\left(\frac{1}{\Omega^2}\right),
\end{equation}
where $\mathcal{H}^n = \frac{1}{T} \int_0^T dt e^{-in\Omega t} \mathcal{H}(t)$. By expanding this commutator and utilizing the spin operator commutation relation $[S_i^\mu, S_j^\nu] = i \epsilon_{\mu\nu\sigma} S_i^\sigma \delta_{ij}$ with $\epsilon_{\mu\nu\sigma}$ the Levi-Civita symbol, one can derive the effective Dzyaloshinskii-Moriya interaction (DMI) between next-nearest-neighbor (NNN) sites
\begin{equation}
	\mathcal{H}_{\text{eff}}^{(1)} = \sum_{\langle\langle\alpha\beta\rangle\rangle} J_{\alpha\beta}^{(1)} \mathbf{S}_\gamma \cdot (\mathbf{S}_\alpha \times \mathbf{S}_\beta),
\end{equation}
where $\gamma$ is the intermediate site acting as the common nearest neighbor connecting the NNN sites $\alpha$ and $\beta$. The coupling strength of the effective DMI is analytically evaluated as $J_{\alpha\beta}^{(1)} = \sum_{n=1}^\infty \frac{1}{n\hbar\Omega}2 J_{n,\perp}^2 \sin(n\Phi_{\alpha\beta})$, where $J_{n,\perp}=J\mathcal{J}_n(\lambda)$, $\mathcal{J}_n$ is the $n$-th order Bessel function of the first kind, and $\Phi_{\alpha\beta}$ is the relative phase difference between the two hopping paths. Because CPL, acting as a uniform field, preserves the spatial inversion $\mathcal{P}$, this term satisfies the symmetry prerequisites for inducing odd-parity spin splitting while concurrently breaking $\mathcal{T}$, as detailed in Supplemental Material (SM)~\cite{SM}. 

We introduce the Holstein-Primakoff transformation in the magnetic ground state and retaining up to quadratic terms:
\begin{equation}
	\begin{aligned}
		&S_{\alpha}^{+} \approx \sqrt{2S}a_{\alpha}, \quad S_{\alpha}^{-} \approx \sqrt{2S}a_{\alpha}^{\dagger}, \quad S_{\alpha}^{z} = S - a_{\alpha}^{\dagger}a_{\alpha} ,\\
		&S_{\beta}^{+} \approx \sqrt{2S}b_{\beta}^{\dagger}, \quad S_{\beta}^{-} \approx \sqrt{2S}b_{\beta}, \quad S_{\beta}^{z} = -S + b_{\beta}^{\dagger}b_{\beta}.
	\end{aligned}
\end{equation}
In the Nambu representation, we use the basis $\Psi_{\mathbf{k}}=\begin{bmatrix}a_{\mathbf{k}}, b_{\mathbf{k}},a_{-\mathbf{k}}^{\dagger}, b_{-\mathbf{k}}^{\dagger}\end{bmatrix}^T$ to construct the matrix $\mathcal{H}=\frac{1}{2}\sum_{\mathbf{k}}\Psi_{\mathbf{k}}^{\dagger}H(\mathbf{k})\Psi_{\mathbf{k}}$. For the first subspace spanned by the basis $\Psi_{1,\mathbf{k}} = [a_{\mathbf{k}}, b_{-\mathbf{k}}^\dagger]^T$, the Hamiltonian reads
\begin{equation}
	H_{1}(\mathbf{k}) = \begin{pmatrix} zJS-D(\mathbf{k}) & \tilde{J}S\gamma_{\mathbf{k}} \\ \tilde{J}S\gamma_{\mathbf{k}}^* & zJS-D(\mathbf{k}) \end{pmatrix}.
\end{equation}
Here, we define the effective exchange constant renormalized by the optical field as $\tilde{J} = J \mathcal{J}_0(\lambda)$, and $z$ is the atomic coordination number. The NN and NNN structure factors are defined as $\gamma_{\mathbf{k}}=\sum_{\boldsymbol{\delta}}e^{i\mathbf{k}\cdot\boldsymbol{\delta}}$ and $D(\mathbf{k}) = 2S^2 \sum_{m=1,3,5} J_{\alpha\beta}^{(1)} \sin(\mathbf{k} \cdot \mathbf{d}_m)$, respectively. The NNN summation is explicitly restricted to three designated NNN vectors $\mathbf{d}_m$ that share an identical relative hopping phase $\Phi$. 
For the second subspace spanned by the basis $\Psi_{2,\mathbf{k}} = [b_{\mathbf{k}}, a_{-\mathbf{k}}^\dagger]^T$, the Hamiltonian is
\begin{equation}
	H_{2}(\mathbf{k}) = \begin{pmatrix} zJS+D(\mathbf{k}) & \tilde{J}S\gamma_{\mathbf{k}}^* \\ \tilde{J}S\gamma_{\mathbf{k}} & zJS+D(\mathbf{k}) \end{pmatrix}.
\end{equation}
Following the Bogoliubov transformation, the energy spectra of the two magnon branches with opposite spin (chirality) are obtained as
\begin{equation}
	\omega_{\uparrow,\downarrow}(\mathbf{k}) = \sqrt{(z {J} S \pm D(\mathbf{k}))^2 - |\tilde{J} S\gamma_{\mathbf{k}}|^2}.
\end{equation}
This dispersion relation strictly satisfies $\omega_{\uparrow}(\mathbf{k}) = \omega_{\downarrow}(-\mathbf{k})$. This provides an analytical verification that that breaking time-reversal symmetry can trigger symmetry-protected odd-parity magnons.

\begin{figure}
	\centering 
	\includegraphics[width=\linewidth]{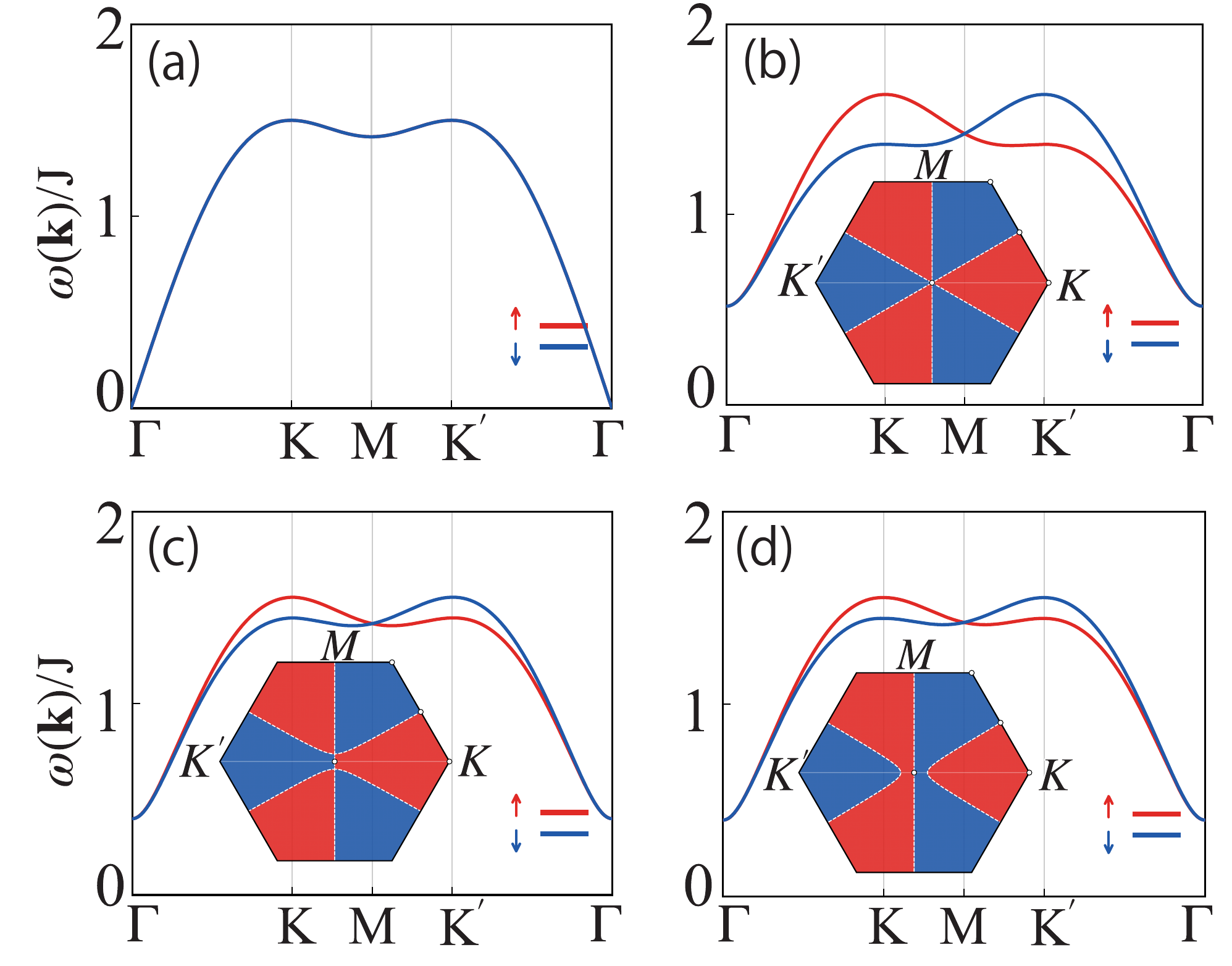}
	\caption{Optically driven odd-parity magnon spin splitting. (a) Magnon band structure in the absence of optical driving, where the two spin/chirality branches are degenerate. (b) Band structure under circularly polarized light driving. The inset highlights the $f$-wave odd-parity characteristics protected by the $C_{3z}$ symmetry. (c), (d) Evolution of the magnon bands under elliptically polarized light, where the reduced symmetry changes the splitting from an $f$-wave to a $p$-wave form. The optical field parameters are $\lambda=0.5$ in (b), $\lambda_x=0.2, \lambda_y=0.5$ in (c), and $\lambda_x=0.5, \lambda_y=0.2$ in (d), with a driving frequency of $\hbar \Omega/J=1$.}
	\label{pic2}
\end{figure}

Figure~\ref{pic2}(b) shows the magnon dispersion relation of the honeycomb lattice antiferromagnet driven by CPL. The time-dependent AC phase introduced by CPL breaks the time-reversal symmetry $\mathcal{T}$, inducing an effective NNN DMI. This nontrivial dynamical modulation lifts the intrinsic spin degeneracy of the system, yielding spin splittings of equal magnitude but opposite signs at the $K$ and $K'$ valleys. Because the system retains the $C_{3z}$ rotational symmetry of the lattice under the CPL field, its momentum-space spin splitting $\Delta\omega(\mathbf{k})$ exhibits an $f$-wave odd-parity profile, as shown in the inset of Fig. \ref{pic2}(b). The system has symmetry $[C_{2\perp}||\mathcal{T}\mathcal{M}_y]$, an operation that performs a mirror reflection across the $y$-axis accompanied by time reversal in real space, while simultaneously flipping the spin in spin space. Dictated by this symmetry, the magnon bands must satisfy the strict constraint $\omega_{\uparrow}(k_x, k_y) = \omega_{\downarrow}(-k_x, k_y)$, compelling the system to maintain exact spin degeneracy along the $\Gamma\text{-M}$ path in momentum space.

To demonstrate the $f$-wave odd-parity nature of the aforementioned spin splitting from an analytical perspective, we expand the low-energy effective dispersion near the $\Gamma$ point. The low-energy form of the effective DMI structure factor $D(\mathbf{k})$ can be analytically expressed as
\begin{equation}
	D(\mathbf{k}) \propto  k_x (k_x^2 - 3k_y^2)+O(k^4).
\end{equation}
This polynomial $k_x (k_x^2 - 3k_y^2)$ perfectly encapsulates the lattice symmetry of the system, in agreement with the basis function of $^{\bar{1}}6/mm^{\bar{1}}m$ listed in Table~\ref{table:odd_parity_2D}. By introducing polar coordinates $(k, \theta)$, the above expression simplifies to $k_x (k_x^2 - 3k_y^2) = k^3 \cos(3\theta)$. Accordingly, the low-energy analytical expression for the spin splitting can be written as $\Delta\omega(\mathbf{k}) \propto 2D(\mathbf{k}) \propto k^3 \cos(3\theta)$. This distinctive $\cos(3\theta)$ angular dependence endows the magnon spin splitting with the symmetry-protected $f$-wave odd-parity characteristics. 

Furthermore, as shown in Figs. \ref{pic2}(c)-(d), when the driving field is switched to elliptically polarized light, the Floquet effective coupling exhibits anisotropy along different spatial directions, thereby breaking the $C_{3z}$ symmetry. Consequently, the momentum-space magnon spin splitting no longer preserves the $f$-wave profile imposed by the lattice $C_{3z}$ symmetry, but instead evolves into a $p$-wave distribution.

This odd-parity mechanism induced by dynamic optical symmetry breaking is broadly applicable. For instance, when this theoretical framework is extended to a square-lattice antiferromagnetic system, the optical field is equally effective in lifting the spin degeneracy. Governed by the $[C_{2\perp}||\mathcal{P}]$ symmetry of the square lattice, the system allows a spin splitting with $p$-wave characteristics (see SM for details)~\cite{SM}. This further corroborates the intrinsic correspondence between static lattice symmetries and dynamic splitting patterns, offering a new route for realizing and manipulating unconventional magnon states across a wide spectrum of antiferromagnetic materials.

\textit{Topological Magnons and Magnon Thermal Hall Eﬀect.}---We now generalize the aforementioned realization mechanism of odd-parity magnons to a bilayer A-type antiferromagnetic system. Typical bilayer A-type antiferromagnets have symmetry $[C_{2\perp} || \mathcal{M}_z]$, where $\mathcal{M}_z$ is the horizontal mirror plane separating the two layers, resulting in spin degeneracy. Therefore, such symmetric bilayer A-type antiferromagnetic systems cannot realize odd-parity spin splitting.

\begin{figure}
	\centering 
	\includegraphics[width=\linewidth]{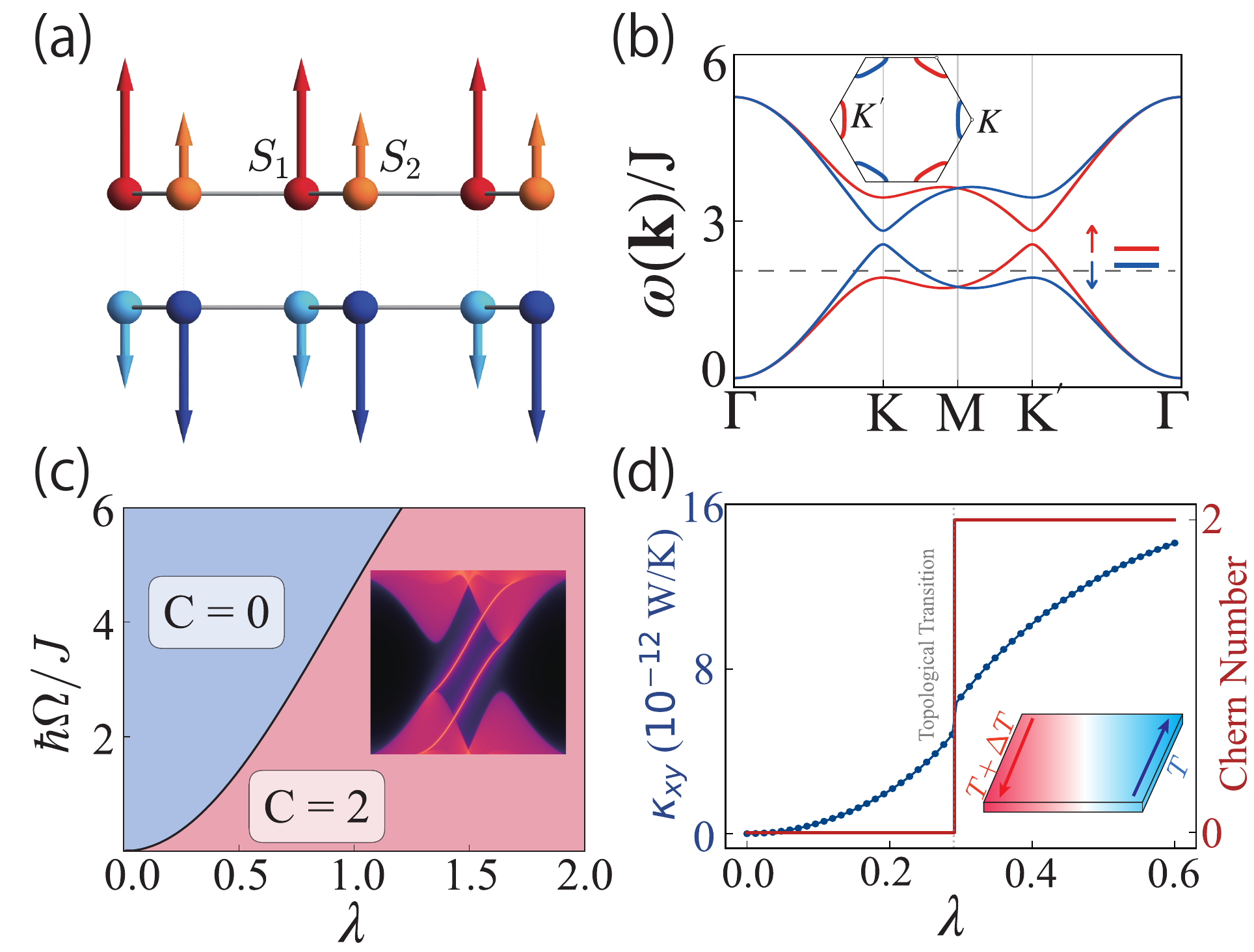}
	\caption{Light-induced topological magnon phase transition and magnon thermal Hall eﬀect. (a) Schematic of the bilayer A-type antiferromagnetic lattice structure featuring spatial asymmetry $S_1 \neq S_2$. (b) $f$-wave magnons under circularly polarized light driving with $\lambda=0.4$ and $\hbar\Omega/J=1$. (c) Topological phase diagram in the plane of the optical driving frequency $\Omega$ and the dimensionless coupling strength $\lambda$, illustrating the system's evolution from a trivial phase $C=0$ to a topological magnon phase $C=2$. The inset shows the corresponding chiral edge states. (d) Evolution of the thermal Hall conductivity $\kappa_{xy}$ (blue dots, left axis) with $J$=1 meV and $T$=50 K and the Chern number (red line, right axis) as a function of $\lambda$. The thermal Hall conductivity exhibits an abrupt jump at the topological phase transition. The common parameters are $S_1=1$, $S_2=0.8$.}
	\label{pic3}
\end{figure}

To overcome this symmetry restriction, one can introduce a spatially asymmetric spin configuration ($S_1 \neq S_2$), as shown in Fig. \ref{pic3}(a), or introduce interlayer sliding in bilayer A-type antiferromagnets. The spin inequivalence breaks the horizontal mirror plane $\mathcal{M}_z$ while strictly preserving $[C_{2\perp} || \mathcal{P}]$, thereby guaranteeing the odd-parity nature of the magnon spectrum. Similarly, we can obtain the magnon energy spectra for A-type bilayer systems with both AA and AB stacking, where AB stacking naturally corresponds to the interlayer sliding (see SM for details). 

When $S_1 \neq S_2$, the magnon spectra strictly satisfy $\omega_{\uparrow}(\mathbf{k}) = \omega_{\downarrow}(-\mathbf{k})$ and exhibit an odd-parity profile, as shown in Fig. \ref{pic3}(b). Around the $K$ and $K'$ valleys, the system can be effectively described by two approximately decoupled Dirac Hamiltonians for the two layers (see SM for details)~\cite{SM}
\begin{equation}
	H_{L_{1,2}}^\eta(\mathbf{q}) = v_F (\eta q_x \sigma_x + q_y \sigma_y) + m_0 \sigma_0 + M_{L_{1,2}}^\eta \sigma_z,
\end{equation}
where $\eta$ is the valley index, the Pauli matrices $\sigma_{x,y,z}$ act on the sublattice pseudospin space, and the constant energy shift is $m_0 = \frac{3J_1(S_1 + S_2)}{2}$. The mass term can be analytically expressed as $M_{L_{1,2}}^\eta = \pm\frac{3J_1(S_1 - S_2)}{2} +\eta S_1S_2J_{\alpha\beta}^{(1)} 3 \sqrt{3}$ with a Semenoff-like trivial mass proportional to $(S_1 - S_2)$ introduced by the spin asymmetry and a Haldane-like topological mass $M_{\text{topo}}$ contributed by the light-induced effective DMI. 

The topological phase transition is driven by the competition between these two mass terms, a process corroborated in Fig. \ref{pic3}. In the static regime without illumination, the system is dominated by the trivial Semenoff mass. Although the spin degeneracy is maintained under the protection of the combinatorial symmetry $[\mathcal{T} || \mathcal{PT}]$, the magnon bands have already opened a trivial gap at the $K$ and $K'$ valleys due to $S_1 \neq S_2$.

Upon the application of circularly polarized light driving, the effective time-reversal symmetry of the system is broken, lifting the spin degeneracy and resulting in $f$-wave splitting. Simultaneously, the light-induced $M_{\text{topo}}$ begins to increase with the dimensionless coupling strength $\lambda$. When the driving reaches a critical threshold of $\lambda \approx 0.3$, where $M_{\text{topo}}$ exactly cancels the trivial Semenoff mass, the band gap at the valleys closes with a pair of Dirac cones of opposite spins (see SM for details)~\cite{SM}. Further increasing the driving, $M_{\text{topo}}$ becomes dominant and reopens a topological gap. The Chern number of upper and lower layers reads $C_{L_{1,2}}=\sum_{\eta}\frac{\eta}{2}\operatorname{sgn}(M^\eta_{L_{1,2}})$, and the total Chern number is $C=C_{L_{1}}+C_{L_{2}}$. The upper and lower layers each contribute a Chern number of $1$, driving the entire system into a topological magnon phase characterized by a total Chern number of $C=2$. This evolution is comprehensively mapped in the topological phase diagram spanned by the driving frequency $\Omega$ and coupling strength $\lambda$, as shown in Fig. \ref{pic3}(c). According to the bulk-boundary correspondence, a topologically nontrivial bulk state inevitably hosts two topologically protected chiral edge states at the boundary, as clearly displayed in the inset of Fig. \ref{pic3}(c).

Driven by an in-plane temperature gradient $\nabla T$, the heat-carrying magnons experience a transverse deflection deflected by the momentum-space Berry curvature $\Omega_{n,z}(\mathbf{k})$, which acts as an effective magnetic field, thereby generating a prominent magnon thermal Hall effect~\cite{katsuraTheoryThermalHall2010a,matsumotoTheoreticalPredictionRotating2011a,onoseObservationMagnonHall2010}. Consequently, the thermal Hall conductivity $\kappa_{xy}$ can serve as a macroscopic observable to characterize the band topology of this odd-parity system. Based on linear response theory, the thermal Hall conductivity of a 2D magnonic system can be evaluated by integrating the Berry curvature over all magnon bands in the Brillouin zone~\cite{matsumotoTheoreticalPredictionRotating2011a}:
\begin{equation}
	\kappa_{xy} = -\frac{k_B^2 T}{\hbar V} \sum_{n, \mathbf{k}} c_2(\rho_n) \Omega_{n,z}(\mathbf{k}),
\end{equation}
where $V$ is the area of the system, and $\rho_n = (e^{\epsilon_{n\mathbf{k}} / k_B T} - 1)^{-1}$ is the Bose-Einstein distribution function. The weight function $c_2(\rho_n) = (1+\rho_n)\left(\ln \frac{1+\rho_n}{\rho_n}\right)^2 - (\ln \rho_n)^2 - 2 \text{Li}_2(-\rho_n)$ decreases monotonically with energy, ensuring that the low-energy magnon excitations at the band bottom dominate the contribution to $\kappa_{xy}$.

In the absence of an external optical field, restricted by the effective time-reversal symmetry, the Berry curvatures from different valleys exactly cancel each other out, yielding a vanishing thermal Hall conductivity. However, when the system develops an odd-parity splitting driven by the CPL—especially as it crosses the critical threshold—the nonzero Berry curvature becomes highly localized around the band gap. This topological evolution is directly verified by the numerical results shown in Fig. \ref{pic3}(d): at the topological phase transition point, accompanying the transition of the total Chern number from a trivial one to $C=2$, the thermal Hall conductivity exhibits a distinct abrupt jump. By measuring the abrupt jump of this thermal transport signal, one can confirm the existence of a topological phase transition and topologically nontrivial phase.

\textit{Material Candidates.}---To verify the feasibility of the aforementioned theoretical framework, we study real 2D van der Waals antiferromagnetic materials. Taking the prototypical monolayer $\text{MnPS}_3$~\cite{leeTunnelingTransportMono2016} as an example, its magnetic $\text{Mn}^{2+}$ ions form a honeycomb lattice, and the ground state exhibits an out-of-plane anisotropic Néel-type collinear antiferromagnetic order. The high-spin state of $S=5/2$ for the $\text{Mn}^{2+}$ ions suppresses quantum spin fluctuations. This inherent lattice and magnetic configuration preserves the spatial inversion symmetry $\mathcal{P}$ centered at the honeycomb hexagon, satisfying the $[C_{2\perp} || \mathcal{P}]$ symmetry protection condition required by our theory. Furthermore, the lattice inherently possesses the $C_{3z}$ rotation symmetry.  For bilayer A-type antiferromagnets, we investigate three representative examples: AB-stacking bilayer $\text{FeBr}_3$~\cite{coleExtremeSensitivityMagnetic2023}, AB'-stacking bilayer $\text{CrI}_3$~\cite{Sivadas2018NL}, and $\text{CrVI}_6$~\cite{panGrowthHighqualityCrI32022}. While all three systems preserve $[C_{2\perp}||\mathcal{P}]$ symmetry, they break $[C_{2\perp}||\mathcal{M}_z]$ symmetry via distinct mechanisms: interlayer sliding in the two bilayers and intrinsic crystalline stacking in $\text{CrVI}_6$.

\begin{figure}
	\centering 
	\includegraphics[width=\linewidth]{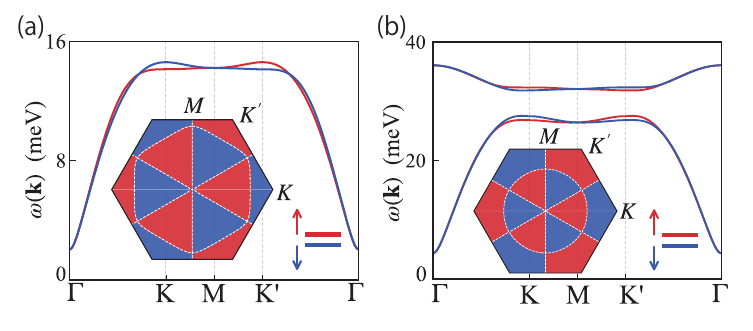}
	\caption{Odd-parity magnon band structures of real van der Waals antiferromagnets. (a), (b) Magnon dispersions of monolayer MnPS$_3$ and bilayer FeBr$_3$ under optical driving with dimensionless coupling strength $\lambda=0.2$ and frequency $\hbar\Omega=1$ meV. The insets show the corresponding $f$-wave spin-splitting patterns. Results for slid bilayer CrI$_3$ and CrVI$_6$ are provided in the Supplemental Material.}
	\label{pic4}
\end{figure}

For a quantitative demonstration, we combine density functional theory with the TB2J tight-binding parameterization method~\cite{heTB2JPythonPackage2021} to extract the  magnetic exchange parameters of these materials, and subsequently calculate their band structures under optical driving. As shown in Fig. \ref{pic4}(a)-(b), upon the application of the optical field, the magnon dispersion exhibits a pronounced spin splitting at the $K$ and $K'$ valleys. Crucially, the broken $C_{3z}$ symmetry in AB'-stacking $\text{CrI}_3$ induces $p$-wave odd-parity splitting, whereas the preserved $C_{3z}$ symmetry in AB-stacking $\text{FeBr}_3$ and $\text{CrVI}_6$ leads to $f$-wave characteristics similar to monolayer $\text{MnPS}_3$ (see SM for details)~\cite{SM}. In real materials, owing to the presence of long-range magnon hopping, the spin splitting in momentum space forms nested inner and outer ring structures with opposite spin polarizations.

\textit{Conclusions and Discussions.}---In summary, we propose odd-parity magnons in collinear magnets, and further tabulate the full spin-point-group classification of the odd-parity magnons including the leading splitting types and splitting forms. It is worth noting that our full spin group classification table for odd-parity splitting applies not only to bosonic systems but also to electronic systems; therefore, it is expected to have broad applicability. To realize odd-parity magnons, we identify a general mechanism, the core of which is breaking the effective time-reversal symmetry $[C_{2\perp}\mathcal{T}||\mathcal{T}]$. Experimentally, this can be achieved by applying CPL or introducing loop currents. CPL not only generates tunable $p$- and $f$-wave magnons in monolayer antiferromagnets, but also drives bilayer antiferromagnets into a topological phase, featuring the emergence of chiral edge states with a Chern number of $C=2$ and an abrupt jump in the magnon thermal Hall conductivity at the topological phase transition. 

Experimentally, polarized inelastic neutron scattering techniques now enable the direct detection and quantitative characterization of the magnon spin-splitting dispersion~\cite{liuChiralSplitMagnon2024,liu2026observation}. We note, however, that relying solely on the bare Aharonov-Casher phase leads to a $1/c^2$-suppressed coupling and would require electric-field strengths beyond current experimental capabilities. More feasible implementations may be achieved through electromagnon- or chiral-phonon-assisted Floquet schemes~\cite{pimenovPossibleEvidenceElectromagnons2006,luoTerahertzControlLinear2025}. In these scenarios, a circularly polarized THz pulse couples directly to spin-dependent electric polarizations or chiral phonons, thereby inducing a tunable dynamical interaction that effectively plays the role of a light-controlled Dzyaloshinskii-Moriya term in the magnon Hamiltonian. This route bypasses the weak bare Aharonov-Casher coupling and can reduce the required driving fields to the experimentally accessible MV/cm regime. Moreover, our framework can also be extended to artificial magnonic lattices, where the relevant spin-point-group symmetries and their dynamical breaking can be engineered with high controllability.

\begin{acknowledgments}
\textit{Acknowledgments.}--- The work is supported by the Science Fund for Creative Research Groups of NSFC (Grant No. 12321004), and the NSF of China (Grant No. 12374055).
\end{acknowledgments}

\textit{Note added.—}
Upon completion of this work, we became aware of~\cite{du2026odd}, which investigates the odd-parity chiral magnons in the relativistic limit and differs from our work, which focuses on the nonrelativistic limit.

\bibliography{oddmagnon}

\appendix

\section*{End matter}

We provide a complete spin-point-group classification of odd-parity magnon spin splitting in 2D Brillouin zones.
This classification assumes that the effective time-reversal symmetry $[C_{2\perp}\mathcal{T}||\mathcal{T}]$ is broken, so that a nonzero odd-parity magnon is no longer symmetry-forbidden.
For each point-group operation $g$, the corresponding spin-point-group element either preserves or flips the magnon spin branch.
The magnon spin splitting
$\Delta\omega(\bm{k})=\omega_{\uparrow}(\bm{k})-\omega_{\downarrow}(\bm{k})$
therefore obeys
\begin{equation}
	\Delta\omega(g\bm{k})=\chi(g)\Delta\omega(\bm{k}),
\end{equation}
where $\chi(g)=+1$ if the spin-point-group element associated with $g$ is spin-preserving, $[E||g]$, and $\chi(g)=-1$ if it is spin-flipping, $[C_{2\perp}||g]$.
Thus, $\Delta\omega(\bm{k})$ transforms as a basis function of a one-dimensional representation $\Gamma_\Delta$ of the underlying point group.
Following this criterion, we systematically examine the one-dimensional irreducible representations of all crystallographic point groups and identify the complete set of collinear spin point groups that can support odd-parity magnon spin splitting in two-dimensional Brillouin zones, as listed in Table~\ref{table:odd_parity_2D}.

When time-reversal symmetry is broken by a Floquet drive, the drive can be viewed, from the symmetry perspective, as generating an effective axial field perpendicular to the two-dimensional plane. Point-group operations that reverse this out-of-plane axial field, such as in-plane twofold rotations or vertical mirror operations, are no longer symmetries by themselves. To maintain symmetries of the Floquet-engineered system, they must be combined with time reversal, namely
\begin{equation}
	g_{\parallel}\rightarrow g_{\parallel}\mathcal{T}.
\end{equation}
Consequently, the symmetry constraint on the magnon spin-splitting function is modified, leading to a different set of allowed odd-parity basis functions. The corresponding Floquet-modified splitting forms are listed in the last column of Table~\ref{table:odd_parity_2D}.

\begin{table*}[b]
	\centering
	\caption{\label{table:odd_parity_2D}Spin-point-group classification of odd-parity magnon spin splitting in two-dimensional Brillouin zones.
		The table lists all collinear spin point groups (SPGs) that can support a nonzero odd-parity magnon spin splitting once the effective time-reversal symmetry $[C_{2\perp}\mathcal{T}||\mathcal{T}]$ is broken.
		Here PG denotes the underlying crystallographic point group, and $\Gamma_\Delta$ denotes the one-dimensional irreducible representation carried by the magnon spin-splitting function $\Delta\omega(\bm{k})=\omega_{\uparrow}(\bm{k})-\omega_{\downarrow}(\bm{k})$.
		A superscript $\bar{1}$ on a spatial generator indicates that this generator is paired with the spin-flipping operation $C_{2\perp}$ in the corresponding collinear spin point group.
		The ``Basic Splitting'' column gives the leading odd-parity basis function after breaking the effective time-reversal symmetry $[C_{2\perp}\mathcal{T}||\mathcal{T}]$, whereas the ``Floquet Splitting'' column gives the corresponding basis function when an additional Floquet-induced out-of-plane effective magnetic field is included. In the basis-function column, $[f_1,f_2]$ means that $f_1$ and $f_2$ furnish equivalent copies of the same one-dimensional representation, so that a general splitting may contain a linear combination $a f_1+b f_2$; by contrast, $[f_1/f_2]$ denotes alternative representative basis functions related by the choice of in-plane coordinate axes, and only one representative applies once the coordinate convention is fixed.
	}
	\footnotesize
	\begin{ruledtabular}
		\begin{tabular}{clll}
			\textbf{PG} & \textbf{SPG ($\Gamma_\Delta$)} & \textbf{Basic Splitting [Function]} & \textbf{Floquet Splitting [Function]}\\
			\colrule
			$\bar{1}$ & ${}^{\bar{1}}\bar{1}$ ($A_u$) & $p$-wave $[x,y]$ & $p$-wave $[x,y]$\\
			$2$       & $^{\bar{1}}2$ ($B$) & $p$-wave $[x/y]$ & $p$-wave $[x/y]$\\
			$2/m$     & $^{\bar{1}}2/m$ ($B_u$) & $p$-wave $[x,y]$ & $p$-wave $[x,y]$\\
			$222$     & $2^{\bar{1}}2^{\bar{1}}2$ ($B_1$) & $p$-wave $[x]$ & $p$-wave $[y]$\\
			& $^{\bar{1}}22^{\bar{1}}2$ ($B_2$) & $p$-wave $[y]$ & $p$-wave $[x]$\\
			$mm2$     & $^{\bar{1}}mm^{\bar{1}}2$ ($B_1$) & $p$-wave $[x]$ & $p$-wave $[y]$ \\
			& $m^{\bar{1}}m^{\bar{1}}2$ ($B_1$) & $p$-wave $[y]$ & $p$-wave $[x]$\\
			$mmm$     & $^{\bar{1}}mmm$ ($B_{1u}$) & $p$-wave $[x]$  & $p$-wave $[y]$\\
			& $m^{\bar{1}}mm$ ($B_{2u}$) & $p$-wave $[y]$  & $p$-wave $[x]$\\
			$\bar{3}$ & $^{\bar{1}}\bar{3}$ ($A_u$) & $f$-wave $[x(x^2-3y^2),y(3x^2-y^2)]$  & $f$-wave $[x(x^2-3y^2),y(3x^2-y^2)]$\\
			$32$      & $3^{\bar{1}}2$ ($A_2$)  & $f$-wave $[x(x^2-3y^2)/y(3x^2-y^2)]$ & $f$-wave $[y(3x^2-y^2)/x(x^2-3y^2)]$\\
			$3m$      & $3^{\bar{1}}2$ ($A_2$)  & $f$-wave $[x(x^2-3y^2)/y(3x^2-y^2)]$ & $f$-wave $[y(3x^2-y^2)/x(x^2-3y^2)]$\\
			$\bar{3}m$& $^{\bar{1}}\bar{3}^{\bar{1}}m$ ($A_{1u}$) & $f$-wave $[x(x^2-3y^2)/y(3x^2-y^2)]$ & $f$-wave $[y(3x^2-y^2)/x(x^2-3y^2)]$\\
			& $^{\bar{1}}\bar{3}m$ ($A_{2u}$) & $f$-wave $[x(x^2-3y^2),y(3x^2-y^2)]$ & $f$-wave $[x(x^2-3y^2),y(3x^2-y^2)]$\\
			$6$       & $^{\bar{1}}6$ ($B$) & $f$-wave $[x(x^2-3y^2),y(3x^2-y^2)]$ & $f$-wave $[x(x^2-3y^2),y(3x^2-y^2)]$\\
			$6/m$     & $^{\bar{1}}6/m$ ($B_u$) & $f$-wave $[x(x^2-3y^2),y(3x^2-y^2)]$ & $f$-wave $[x(x^2-3y^2),y(3x^2-y^2)]$\\
			$622$     & $^{\bar{1}}62^{\bar{1}}2$ ($B_1$) & $f$-wave $[y(3x^2-y^2)]$ & $f$-wave $[x(x^2-3y^2)]$\\
			& $^{\bar{1}}6^{\bar{1}}22$ ($B_2$) & $f$-wave $[x(x^2-3y^2)]$ & $f$-wave $[y(3x^2-y^2)]$ \\
			$6mm$     & $^{\bar{1}}6m^{\bar{1}}m$ ($B_1$) & $f$-wave $[y(3x^2-y^2)]$ & $f$-wave $[x(x^2-3y^2)]$\\
			& $^{\bar{1}}6^{\bar{1}}mm$ ($B_2$) & $f$-wave $[x(x^2-3y^2)]$ & $f$-wave $[y(3x^2-y^2)]$\\
			$\bar{6}m2$&$\bar{6}^{\bar{1}}m^{\bar{1}}2$ ($A_2'$) & $f$-wave $[x(x^2-3y^2)]$& $f$-wave $[y(3x^2-y^2)]$ \\
			$6/mmm$   & $^{\bar{1}}6/mm^{\bar{1}}m$ ($B_{1u}$) & $f$-wave $[y(3x^2-y^2)]$ & $f$-wave $[x(x^2-3y^2)]$\\
			& $^{\bar{1}}6/m^{\bar{1}}mm$ ($B_{2u}$) & $f$-wave $[x(x^2-3y^2)]$& $f$-wave $[y(3x^2-y^2)]$
		\end{tabular}
	\end{ruledtabular}
\end{table*}

\end{document}


\title{Supplementary Material for  "Odd-Parity Magnons"}
	\author{Pu Zhang}
	\author{Sun-Bo Xie}
	\author{Junxi Yu}
	\author{Yichen Liu}
	\author{Cheng-Cheng Liu}
	\email{ccliu@bit.edu.cn}
	\affiliation{Centre for Quantum Physics, Key Laboratory of Advanced Optoelectronic Quantum Architecture and Measurement (MOE), School of Physics, Beijing Institute of Technology, Beijing 100081, China}
	\maketitle
	\tableofcontents
	
	\section{Aharonov-Casher Effect}
	
	In the absence of a macroscopic magnetic field ($\pmb{B}=0$) but in the presence of an applied electric field $\pmb{E}$, the Aharonov-Casher (AC) phase of a neutral particle with an intrinsic magnetic moment $\mu$ can be derived from the non-relativistic limit of the Dirac-Pauli Hamiltonian. In the standard Dirac representation, the Hamiltonian is written in the full $4 \times 4$ block matrix form
	\begin{equation}
		H = \begin{pmatrix} M c^2 & c\pmb{\sigma} \cdot \left(\pmb{p} - i\frac{\mu}{c^2}\pmb{E}\right) \\ c\pmb{\sigma} \cdot \left(\pmb{p} + i\frac{\mu}{c^2}\pmb{E}\right) & -M c^2 \end{pmatrix},
	\end{equation}
	where $M$ is the rest mass of the particle, and $\pmb{\sigma}$ are the Pauli matrices. The four-component spinor wave function can be decomposed into a large component $u$ and a small component $v$. Substituting this into the stationary Schrödinger equation and separating the variables yields the coupled equations 
	\begin{align}
		(M c^2 + E_{NR})u &= M c^2 u + c\pmb{\sigma} \cdot \left(\pmb{p} - i\frac{\mu}{c^2}\pmb{E}\right)v \label{8},\\
		(M c^2 + E_{NR})v &= c\pmb{\sigma} \cdot \left(\pmb{p} + i\frac{\mu}{c^2}\pmb{E}\right)u - M c^2 v \label{9}.
	\end{align}
	In the non-relativistic limit, the non-relativistic kinetic energy $E_{NR}$ of the system is much smaller than the rest mass energy, i.e., $E_{NR} \ll 2M c^2$. In this case, $2M c^2 + E_{NR}$ can be approximated as $2M c^2$. Thus, we can approximate the small component $v$ as a function of the large component $u$
	\begin{equation}
		v \approx \frac{1}{2Mc} \pmb{\sigma} \cdot \left(\pmb{p} + i\frac{\mu}{c^2}\pmb{E}\right) u 
		\label{10}.
	\end{equation}
	Substituting this back into Eq.~\ref{8} for the large component $u$, the $Mc^2 u$ terms on both sides cancel out, and we obtain the low-energy effective Schrödinger equation describing only the positive-energy states, $E_{NR} u = H_{NR} u$, where the effective Hamiltonian is
	\begin{equation}
		H_{NR} = \frac{1}{2M} \left[ \pmb{\sigma} \cdot \left(\pmb{p} - i\frac{\mu}{c^2}\pmb{E}\right) \right] \left[ \pmb{\sigma} \cdot \left(\pmb{p} + i\frac{\mu}{c^2}\pmb{E}\right) \right].
		\label{11}
	\end{equation}
	Expanding the operator product, the effective Hamiltonian becomes 
	\begin{equation}
		H_{NR} = \frac{1}{2M} \left[ p^2 + \left(\frac{\mu E}{c^2}\right)^2 - \frac{2\mu}{c^2} \pmb{\sigma} \cdot (\pmb{p} \times \pmb{E}) \right].
	\end{equation}
	Using the identity $\pmb{\sigma} \cdot (\pmb{p} \times \pmb{E}) = \pmb{p} \cdot (\pmb{E} \times \pmb{\sigma})$, and defining the intrinsic magnetic moment vector operator $\boldsymbol{\mu} = \mu\pmb{\sigma}$, this term can be rewritten as $-\frac{2}{c^2} \pmb{p} \cdot (\pmb{E} \times \boldsymbol{\mu})$. To clarify the gauge structure, we complete the square for the momentum squared term and the spin-orbit coupling term 
	\begin{equation}
		p^2 - \frac{2}{c^2} \pmb{p} \cdot (\pmb{E} \times \boldsymbol{\mu}) = \left( \pmb{p} - \frac{1}{c^2} \pmb{E} \times \boldsymbol{\mu} \right)^2 - \left( \frac{1}{c^2} \pmb{E} \times \boldsymbol{\mu} \right)^2.
	\end{equation}
	This result establishes the minimal coupling form between the intrinsic magnetic moment and the electric field, where the induced equivalent gauge vector potential is 
	\begin{equation}
		\mathbf{A}_{AC} = \frac{1}{c^2} \pmb{E} \times \boldsymbol{\mu}.
	\end{equation}
	In lattice dynamics, the gauge vector potential $\mathbf{A}_{AC}$ in continuous space is introduced into the hopping integral of the tight-binding model through the Peierls substitution~\cite{aharonovTopologicalQuantumEffects1984a} 
	\begin{equation}
		t_{ij} \rightarrow t_{ij} \exp\left( \frac{i}{\hbar} \int_{\mathbf{r}_i}^{\mathbf{r}_j} \mathbf{A}_{AC} \cdot d\mathbf{l}\right).
	\end{equation}
	
	\section{Effective Floquet antiferromagnetic Heisenberg model}
	
	The Heisenberg Hamiltonian containing nearest-neighbor exchange interactions reads
	\begin{equation}
		\mathcal{H}_0 = J \sum_{\langle \alpha,\beta \rangle} \mathbf{S}_\alpha \cdot \mathbf{S}_\beta+K \sum_{\alpha} (S_\alpha^z)^2,
	\end{equation}
	where $K<0$ represents the easy-axis anisotropy, which stabilizes the N\'eel vector along the $\hat{z}$ direction. For simplicity, we will ignore this term in the following analysis. Since adjacent spins in an antiferromagnetic system tend to align antiparallel to minimize the energy ($J>0$), we introduce a two-sublattice model comprising sublattices A and B.
	
	Here, $\mathbf{S}_\alpha$ and $\mathbf{S}_\beta$ denote the magnetic spin vector operators located at the lattice coordinates $\mathbf{r}_\alpha$ and $\mathbf{r}_\beta$, respectively 
	\begin{equation}
		\mathbf{S}_\alpha \cdot \mathbf{S}_\beta = S_\alpha^x S_\beta^x + S_\alpha^y S_\beta^y + S_\alpha^z S_\beta^z.
	\end{equation}
	
	Using the relations $S^\pm = S^x \pm iS^y$, the inverse transformations are 
	\begin{equation}
		S^x = \frac{1}{2}(S^+ + S^-),\quad S^y = \frac{1}{2i}(S^+ - S^-).
	\end{equation}
	Substituting these into the product of the $x$ and $y$ components gives 
	\begin{equation}
		\begin{aligned}
			S_\alpha^x S_\beta^x + S_\alpha^y S_\beta^y &= \frac{1}{4}(S_\alpha^+ + S_\alpha^-)(S_\beta^+ + S_\beta^-) - \frac{1}{4}(S_\alpha^+ - S_\alpha^-)(S_\beta^+ - S_\beta^-)\\
			&= \frac{1}{2} (S_\alpha^+ S_\beta^- + S_\alpha^- S_\beta^+).
		\end{aligned}
	\end{equation}
	
	Therefore, the Heisenberg model can be rewritten as 
	\begin{equation}
		\mathcal{H} = \sum_{\langle\alpha,\beta\rangle} \left[ \frac{J}{2} \left( S_\alpha^- S_\beta^+  + \text{h.c.} \right) + J S_\alpha^z S_\beta^z \right].
	\end{equation}
	When a magnetic material placed in the $xy$ plane is irradiated normally by a circularly polarized laser (electromagnetic field), its dominant component is an oscillating electric field $\mathbf{E}(t)$. In this electric field background, magnon quasiparticles undergoing hopping transitions will acquire a time-dependent AC phase due to their magnetic dipole moments.
	
	\begin{equation}
		\theta_{\alpha\beta}(t) = \frac{g\mu_B}{\hbar c^2} \int_{r_\alpha}^{r_\beta} \left( \mathbf{E}(t) \times \hat{z} \right) \cdot d\boldsymbol{\ell},
	\end{equation}
	where $\hbar$ is the reduced Planck constant and $c$ is the speed of light. The oscillating electric field satisfies $\mathbf{E}(t) = -\partial \mathbf{A}(t)/\partial t$, where $\mathbf{A}(t)$ is a time-periodic vector potential. We choose the oscillating electric field to have the following form 
	\begin{equation}
		\mathbf{E}(t) \times \hat{z} = E_0\left(\sin\Omega t, -\cos\Omega t, 0\right),
	\end{equation}
	where $E_0$ is the electric field amplitude. The corresponding time-dependent Hamiltonian is 
	\begin{equation}
		\mathcal{H}(t) = \sum_{\langle\alpha,\beta\rangle} \left[ \frac{J}{2} \left( S_\alpha^- S_\beta^+ e^{i\theta_{\alpha\beta}(t)} + \text{h.c.} \right) + J S_\alpha^z S_\beta^z \right].
	\end{equation}
	
	By defining a polar angle $\phi_{\alpha\beta}$ corresponding to the vector direction from lattice site $\pmb{r}_{\alpha}$ to $\pmb{r}_{\beta}$, one can deduce that $\theta_{\alpha\beta}(t) = \lambda \sin(\Omega t - \phi_{\alpha\beta})$, where
	\begin{equation}
		\lambda = \frac{g\mu_B E_0 r_{\alpha\beta}}{\hbar c^2},
	\end{equation}
	and $r_{\alpha\beta}$ is the distance between the two lattice sites. The static, time-independent effective Hamiltonian can be expanded in a power series of $\Omega^{-1}$ as $\mathcal{H}_{\text{eff}} \approx \mathcal{H}_{\text{eff}}^{(0)} + \mathcal{H}_{\text{eff}}^{(1)}$. The zeroth-order term is $\mathcal{H}_{\text{eff}}^{(0)} = \mathcal{H}^0$, and the first-order correction term is 
	\begin{equation}
		\mathcal{H}_{\text{eff}}^{(1)} = \sum_{n=1}^{\infty} \frac{1}{n \Omega}[\mathcal{H}^n, \mathcal{H}^{-n}],
	\end{equation}
	where $\mathcal{H}^n = \frac{1}{T} \int_0^T dt e^{-in\Omega t} \mathcal{H}(t)$.
	
	The specific form of the corresponding Fourier components is 
	\begin{equation}
		\mathcal{H}^n = -\sum_{\langle\alpha\beta\rangle} \left[ \frac{J_{n,\perp}}{2} \left( S_\alpha^- S_\beta^+ e^{-in\phi_{\alpha\beta}} + (-1)^nS_\alpha^+ S_\beta^- e^{-in\phi_{\alpha\beta}} \right) + J\delta_{n,0} S_\alpha^z S_\beta^z \right],
	\end{equation}
	where $J_{n,\perp} = J \mathcal{J}_n(\lambda)$, and $\mathcal{J}_n(\lambda)$ is the Bessel function of integer order $n$; the Kronecker $\delta$ function $\delta_{n,0}$ equals 1 when $n = 0$ and 0 otherwise. In this derivation, we utilized the mathematical relations 
	\begin{equation}
		e^{iz\sin(x)} = \sum_{n=-\infty}^{\infty} \mathcal{J}_n(z) e^{inx}, \quad \mathcal{J}_n(-z)=(-1)^n\mathcal{J}_n(z).
	\end{equation}
	The zeroth-order effective Hamiltonian takes the form $\mathcal{H}_{\text{eff}}^{(0)}=\mathcal{H}^0$ with the explicit expression 
	\begin{equation}
		\mathcal {H}_{\text{eff}}^{(0)}=\sum _{\langle \alpha \beta \rangle }\big [J_{0,\perp }(S_{\alpha }^{x}S_{\beta }^{x}+S_{\alpha }^{y}S_{\beta }^{y})+JS_{\alpha }^{z}S_{\beta }^{z}\big ],
	\end{equation}
	where $J_{0, \perp}=J \mathcal{J}_{0}(\lambda)$. Thus, the zeroth-order term yields an XXZ-type Hamiltonian.
	
	The expression for the first-order correction term of the effective Hamiltonian is 
	\begin{equation}
		\mathcal{H}_{\text{eff}}^{(1)}=\sum_{n=1}^{\infty} \frac{1}{n\Omega}\left[\mathcal{H}^{n}, \mathcal{H}^{-n}\right].
	\end{equation} 
	Since the longitudinal term is time-independent, its Fourier components strictly vanish for $n \neq 0$. Thus, for $n \ge 1$, $\mathcal{H}^n$ only contains the transverse components and can be explicitly written as
	\begin{equation}
		\mathcal{H}^n = -\sum_{\langle \alpha\beta \rangle} \frac{J_{n,\perp}}{2} \left( S_\alpha^- S_\beta^+ e^{-in\phi_{\alpha\beta}} + (-1)^n S_\alpha^+ S_\beta^- e^{-in\phi_{\alpha\beta}} \right).
	\end{equation}
	To calculate the first-order correction, we need to evaluate $\mathcal{H}^{-n}$. By substituting $-n$ for $n$ in the above equation and utilizing the symmetry property of Bessel functions $J_{-n,\perp} = J \mathcal{J}_{-n}(\lambda) = (-1)^n J_{n,\perp}$ 
	\begin{equation}
		\mathcal{H}^{-n} = -\sum_{\langle \rho\gamma \rangle} \frac{J_{n,\perp}}{2} \left( (-1)^n S_\rho^- S_\gamma^+ e^{in\phi_{\rho\gamma}} + S_\rho^+ S_\gamma^- e^{in\phi_{\rho\gamma}} \right).
	\end{equation}
	
	The first-order correction formula is $\mathcal{H}_{eff}^{(1)} = \sum_{n=1}^\infty \frac{1}{n\Omega} [\mathcal{H}^n, \mathcal{H}^{-n}]$. By expanding and multiplying $\mathcal{H}^n$ and $\mathcal{H}^{-n}$, we focus on the cross terms that generate three-spin interactions. Here, we use the commutation relation 
	\begin{equation}
		[S_\alpha^+ S_\beta^-, S_\rho^+ S_\gamma^-] = -2(\delta_{\beta\rho} S_\beta^z S_\alpha^+ S_\gamma^- - \delta_{\alpha\gamma} S_\alpha^z S_\rho^+ S_\beta^-).
	\end{equation}
	Let $\gamma$ be the intermediate lattice site connecting the next-nearest-neighbor sites $\alpha$ and $\beta$, meaning we treat $\langle \alpha, \gamma \rangle$ and $\langle \gamma, \beta \rangle$ as nearest-neighbor bonds. We extract the term containing $S_\alpha^+ S_\gamma^-$ from $\mathcal{H}^n$ and the term containing $S_\gamma^+ S_\beta^-$ from $\mathcal{H}^{-n}$, then compute their commutator 
	\begin{equation}
		\left[ -\frac{J_{n,\perp}}{2} (-1)^n S_\alpha^+ S_\gamma^- e^{-in\phi_{\alpha\gamma}}, -\frac{J_{n,\perp}}{2} S_\gamma^+ S_\beta^- e^{in\phi_{\gamma\beta}} \right] = -\frac{J_{n,\perp}^2}{4} (-1)^n e^{-in\phi_{\alpha\gamma}} e^{in\phi_{\gamma\beta}} [S_\alpha^+ S_\gamma^-, S_\gamma^+ S_\beta^-].
	\end{equation}
	Substituting the commutation relation $[S_\alpha^+ S_\gamma^-, S_\gamma^+ S_\beta^-] = -2 S_\gamma^z S_\alpha^+ S_\beta^-$, the result is 
	\begin{equation}
		-\frac{J_{n,\perp}^2}{2} (-1)^n e^{-in\phi_{\alpha\gamma} + in\phi_{\gamma\beta}} S_\gamma^z S_\alpha^+ S_\beta^-.
	\end{equation}
	Similarly, computing the commutator for the cross terms containing $S_\alpha^- S_\gamma^+$ and $S_\gamma^- S_\beta^+$, and utilizing $[S_\alpha^- S_\gamma^+, S_\gamma^- S_\beta^+] = 2 S_\gamma^z S_\alpha^- S_\beta^+$ yields
	\begin{equation}
		\frac{J_{n,\perp}^2}{2} (-1)^n e^{-in\phi_{\alpha\gamma} + in\phi_{\gamma\beta}} S_\gamma^z S_\alpha^- S_\beta^+.
	\end{equation}
	
	Combining the above two terms and factoring out common terms 
	\begin{equation}
		-\frac{J_{n,\perp}^2}{2} (-1)^n e^{-in\phi_{\alpha\gamma} + in\phi_{\gamma\beta}} S_\gamma^z (S_\alpha^+ S_\beta^- - S_\alpha^- S_\beta^+).
	\end{equation}
	Using the expansion of spin raising/lowering operators and transverse components $S^\pm = S^x \pm iS^y$, one can show that the term inside the parentheses is equivalent to the $z$-component of a cross product 
	\begin{equation}
		S_\alpha^+ S_\beta^- - S_\alpha^- S_\beta^+ = -2i (S_\alpha^x S_\beta^y - S_\alpha^y S_\beta^x) = -2i (\pmb{S}_\alpha \times \pmb{S}_\beta)_z.
	\end{equation}
	Therefore, the spin part simplifies to a three-spin chiral term  $-2i \pmb{S}_\gamma \cdot (\pmb{S}_\alpha \times \pmb{S}_\beta)$. In a bipartite lattice, the next-nearest-neighbor sites $\alpha$ and $\beta$ belong to the same sublattice, while the intermediate site $\gamma$ belongs to the opposite sublattice. Next, we handle the phase part. We define $\Phi_{\alpha\beta} = \phi_{\alpha\gamma} - \phi_{\beta\gamma}$. Since $\phi_{\gamma\beta}$ and $\phi_{\beta\gamma}$ are vector angles pointing in opposite directions, they satisfy $\phi_{\gamma\beta} = \phi_{\beta\gamma} \pm \pi$, hence $e^{in\phi_{\gamma\beta}} = (-1)^n e^{in\phi_{\beta\gamma}}$. Substituting this into the exponential term 
	\begin{equation}
		e^{-in\phi_{\alpha\gamma}} e^{in\phi_{\gamma\beta}} = e^{-in\phi_{\alpha\gamma}} (-1)^n e^{in\phi_{\beta\gamma}} = (-1)^n e^{-in(\phi_{\alpha\gamma} - \phi_{\beta\gamma})} = (-1)^n e^{-in\Phi_{\alpha\beta}}.
	\end{equation}
	Multiplying this with the original $(-1)^n$ in the coefficient yields $(-1)^n \times (-1)^n = 1$. Consequently, the contribution from each next-nearest-neighbor path $\langle  \alpha, \gamma, \beta \rangle$ simplifies to 
	\begin{equation}
		i J_{n,\perp}^2 e^{-in\Phi_{\alpha\beta}} \pmb{S}_\gamma \cdot (\pmb{S}_\alpha \times \pmb{S}_\beta).
	\end{equation}
	
	After substituting the above equation into the first-order correction formula $\mathcal{H}_{eff}^{(1)} = \sum_{n=1}^\infty \frac{1}{n\Omega} [\mathcal{H}^n, \mathcal{H}^{-n}]$ and expanding the complex exponential via Euler's formula $e^{-in\Phi_{\alpha\beta}} = \cos(n\Phi_{\alpha\beta}) - i\sin(n\Phi_{\alpha\beta})$, we sum over the triangular clusters of the entire lattice, since the Hamiltonian is Hermitian and the cross product obeys antisymmetry $\pmb{S}_\beta \times \pmb{S}_\alpha = -\pmb{S}_\alpha \times \pmb{S}_\beta$ along with $\Phi_{\beta\alpha} = -\Phi_{\alpha\beta}$, the real cosine part will mutually cancel during the exchange summation of $\alpha$ and $\beta$, leaving only the sine part resulting from the broken time-reversal symmetry. The final form of the synthetic effective Hamiltonian is 
	\begin{equation}
		\mathcal{H}_{eff}^{(1)} = \sum_{\langle\langle\alpha\beta\rangle\rangle} J_{\alpha\beta}^{(1)} \mathbf{S}_\gamma \cdot (\mathbf{S}_\alpha \times \mathbf{S}_\beta).
	\end{equation}
The first-order Floquet correction generates a three-spin scalar-chirality term, which reduces to an effective next-nearest-neighbor Dzyaloshinskii-Moriya interaction (DMI) after projecting the intermediate spin onto the collinear magnetic background with strength
	\begin{equation}
		J_{\alpha\beta}^{(1)} = \sum_{n=1}^\infty \frac{1}{n\Omega}2 J_{n,\perp}^2 \sin(n\Phi_{\alpha\beta}).
	\end{equation}
	
	The total effective Hamiltonian consists of the zeroth-order term and the first-order correction term  $\mathcal{H}_{eff} = \mathcal{H}_{eff}^{(0)} + \mathcal{H}_{eff}^{(1)}$.
	
	\section{Holstein-Primakoff Transformation}
	
	In an antiferromagnetic system, the excitation modes of sublattice A and sublattice B are opposite. For the spin-up sublattice A, we introduce the bosonic operator $a$. For sublattice B with spin down, the ground state corresponds to the minimum value $-S$, hence the action of the raising and lowering operators for sublattice B is inverted, denoted by $b$. The antiferromagnetic Holstein-Primakoff (HP) transformation formulas are as follows.
	
	Sublattice A, lattice site $\alpha$  
	\begin{equation}
		S_{\alpha}^{+}\approx\sqrt{2S}a_{\alpha},\quad S_{\alpha}^{-}\approx\sqrt{2S}a_{\alpha}^{\dagger},\quad S_{\alpha}^{z}=S-a_{\alpha}^{\dagger}a_{\alpha},
	\end{equation}
	
	Sublattice B, lattice site $\beta$  
	\begin{equation}
		S_{\beta}^{+}\approx\sqrt{2S}b_{\beta}^{\dagger},\quad
		S_{\beta}^{-}\approx\sqrt{2S}b_{\beta},\quad
		S_{\beta}^{z}=-S+b_{\beta}^{\dagger}b_{\beta},
	\end{equation}
	
	Considering the nearest-neighbor interactions $\alpha \in A, \beta \in B$, the longitudinal term is 
	\begin{equation}
		S_{\alpha}^{z}S_{\beta}^{z}=(S-a_{\alpha}^{\dagger}a_{\alpha})(-S+b_{\beta}^{\dagger}b_{\beta})\approx -S^{2}+S(a_{\alpha}^{\dagger}a_{\alpha}+b_{\beta}^{\dagger}b_{\beta}),
	\end{equation}
	where we neglect higher-order four-operator terms.
	
	For the transverse components, since $S_{\beta}^{+}$ corresponds to the creation operator $b_{\beta}^{\dagger}$, two-magnon excitations are generated here 
	\begin{equation}
		\begin{aligned}
			S_{\alpha}^{x}S_{\beta}^{x}+S_{\alpha}^{y}S_{\beta}^{y}&=\frac{1}{2}(S_{\alpha}^{+}S_{\beta}^{-}+S_{\alpha}^{-}S_{\beta}^{+})\\
			&\approx\frac{1}{2}(\sqrt{2S}a_{\alpha}\sqrt{2S}b_{\beta}+\sqrt{2S}a_{\alpha}^{\dagger}\sqrt{2S}b_{\beta}^{\dagger})\\
			&=S(a_{\alpha}b_{\beta}+a_{\alpha}^{\dagger}b_{\beta}^{\dagger}).
		\end{aligned}
	\end{equation}
	
	Substituting these into $\mathcal{H}_{eff}^{(0)}$ and omitting the constant term $JS^2$ 
	\begin{equation}
		\mathcal{H}_{eff}^{(0)}=\sum_{\langle\alpha\beta\rangle}[JS(a_{\alpha}^{\dagger}a_{\alpha}+b_{\beta}^{\dagger}b_{\beta})+J_{0,\perp}S(a_{\alpha}b_{\beta}+a_{\alpha}^{\dagger}b_{\beta}^{\dagger})].
	\end{equation}
	
	The first-order correction term is summed over next-nearest-neighbor lattice sites $\langle\langle\alpha\beta\rangle\rangle$. In a two-sublattice setup, next-nearest neighbors reside within the same sublattice. We need to separately compute the DMI within sublattice A and sublattice B. For next-nearest neighbors within sublattice A ($\alpha, \beta \in A$), the intermediate site is $\gamma \in B$. Because $\gamma \in B$, its classical expectation value is $S_{\gamma}^{z}\approx -S$. 
	The $z$-component of the cross product is 
	\begin{equation}
		(\pmb{S}_{\alpha}\times\pmb{S}_{\beta})_{z}=\frac{i}{2}(S_{\alpha}^{+}S_{\beta}^{-}-S_{\alpha}^{-}S_{\beta}^{+}),
	\end{equation}
	Substituting the HP transformation for lattice A 
	\begin{equation}
		\frac{i}{2}(2Sa_{\alpha}a_{\beta}^{\dagger}-2Sa_{\alpha}^{\dagger}a_{\beta})=iS(a_{\beta}^{\dagger}a_{\alpha}-a_{\alpha}^{\dagger}a_{\beta}),
	\end{equation}
	Multiplying by $S_{\gamma}^{z}$ 
	\begin{equation}
		\mathcal{H}_{eff, A}^{(1)}=\sum_{\langle\langle\alpha\beta\rangle\rangle\in A}-iJ_{\alpha\beta}^{(1)}S^{2}(a_{\beta}^{\dagger}a_{\alpha}-a_{\alpha}^{\dagger}a_{\beta}).
	\end{equation}
	
	For next-nearest neighbors within sublattice B ($\alpha, \beta \in B$), the intermediate site is $\gamma \in A$. Because $\gamma \in A$, its classical expectation value is $S_{\gamma}^{z}\approx S$. 
	Substituting the HP transformation for lattice B 
	\begin{equation}
		\begin{aligned}
			(\pmb{S}_{\alpha}\times\pmb{S}_{\beta})_{z}&=\frac{i}{2}(\sqrt{2S}b_{\alpha}^{\dagger}\sqrt{2S}b_{\beta}-\sqrt{2S}b_{\alpha}\sqrt{2S}b_{\beta}^{\dagger})\\
			&=iS(b_{\alpha}^{\dagger}b_{\beta}-b_{\alpha}b_{\beta}^{\dagger}).
		\end{aligned}
	\end{equation}
	Using the commutation properties of bosons at lattice sites ($b_{\alpha}b_{\beta}^{\dagger}=b_{\beta}^{\dagger}b_{\alpha}$) 
	\begin{equation}
		iS(b_{\alpha}^{\dagger}b_{\beta}-b_{\beta}^{\dagger}b_{\alpha}) = -iS(b_{\beta}^{\dagger}b_{\alpha}-b_{\alpha}^{\dagger}b_{\beta}),
	\end{equation}
	Multiplying by $S_{\gamma}^{z} \approx S$ 
	\begin{equation}
		\mathcal{H}_{eff, B}^{(1)}=\sum_{\langle\langle\alpha\beta\rangle\rangle\in B}-iJ_{\alpha\beta}^{(1)}S^{2}(b_{\beta}^{\dagger}b_{\alpha}-b_{\alpha}^{\dagger}b_{\beta}).
	\end{equation}
	
	Following the HP transformation, the Floquet DMI terms on sublattices A and B in the antiferromagnetic system acquire the identical mathematical form, but due to the differing definitions of the HP transformations for the two sublattices, the DMI effects on sublattices A and B have opposite signs.
	Thus, we obtain the effective Hamiltonian in terms of the raising and lowering operators for two sublattices
	\begin{equation}\label{eq H_eff_ab}
		\begin{aligned}
			\mathcal{H}_{eff} =& \mathcal{H}_{eff}^{(0)} + \mathcal{H}_{eff,A}^{(1)} + \mathcal{H}_{eff,B}^{(1)} \\
			= & \sum_{\langle\alpha\beta\rangle}[JS(a_{\alpha}^{\dagger}a_{\alpha}+b_{\beta}^{\dagger}b_{\beta})-J_{0,\perp}S(a_{\alpha}b_{\beta}+a_{\alpha}^{\dagger}b_{\beta}^{\dagger})]-iS^{2}\sum_{\langle\langle\alpha\beta\rangle\rangle}J_{\alpha\beta}^{(1)}(a_{\beta}^{\dagger}a_{\alpha}-a_{\alpha}^{\dagger}a_{\beta}+b_{\beta}^{\dagger}b_{\alpha}-b_{\alpha}^{\dagger}b_{\beta}) \\
			= & zJS\sum_{\alpha}(a_{\alpha}^{\dagger}a_{\alpha}+b_{\alpha}^{\dagger}b_{\alpha})-J_{0,\perp}S\sum_{\langle\alpha\beta\rangle}(a_{\alpha}b_{\beta}+a_{\alpha}^{\dagger}b_{\beta}^{\dagger})
			-iS^{2}\sum_{\langle\langle\alpha\beta\rangle\rangle}J_{\alpha\beta}^{(1)}(a_{\beta}^{\dagger}a_{\alpha}-a_{\alpha}^{\dagger}a_{\beta}+b_{\beta}^{\dagger}b_{\alpha}-b_{\alpha}^{\dagger}b_{\beta}).
		\end{aligned}
	\end{equation}
	The diagonal term $\sum_{\langle\alpha\beta\rangle}a_{\alpha}^{\dagger}a_{\alpha}$ is calculated $z$ times for each A site, where $z$ is the coordination number.
	
	\section{Nambu Representation}
	
	To transform the operators from real space $\pmb{r}$ to momentum space $\pmb{k}$, the Fourier transforms are given by 
	\begin{equation}
		\begin{aligned}
			a_{\alpha}&=\sqrt{\frac{2}{N}}\sum_{\pmb{k}}e^{i\pmb{k}\cdot\pmb{r}_{\alpha}}a_{\pmb{k}}, \quad a_{\alpha}^{\dagger}=\sqrt{\frac{2}{N}}\sum_{\pmb{k}}e^{-i\pmb{k}\cdot\pmb{r}_{\alpha}}a_{\pmb{k}}^{\dagger},\\
			b_{\alpha}&=\sqrt{\frac{2}{N}}\sum_{\pmb{k}}e^{i\pmb{k}\cdot\pmb{r}_{\alpha}}b_{\pmb{k}}, \quad b_{\alpha}^{\dagger}=\sqrt{\frac{2}{N}}\sum_{\pmb{k}}e^{-i\pmb{k}\cdot\pmb{r}_{\alpha}}b_{\pmb{k}}^{\dagger}.\\
		\end{aligned}
	\end{equation}
	
	We process the real-space Hamiltonian Eq.~(\ref{eq H_eff_ab}) term by term. For the on-site terms 
	\begin{equation}
		\begin{aligned}
			\sum_{\alpha} a^{\dagger}_{\alpha}a_{\alpha} &= \sum_{\alpha} \frac{2}{N} \sum_{\pmb{k}}\sum_{\pmb{k}^{'}}e^{i(\pmb{k}-\pmb{k}^{'})\cdot \pmb{r}_{\alpha}}a^{\dagger}_{\pmb{k}^{'}}a_{\pmb{k}} \\
			&=\sum_{\pmb{k}}a^{\dagger}_{\pmb{k}}a_{\pmb{k}},
		\end{aligned}
	\end{equation}
	Similarly, $\sum_{\alpha}b^{\dagger}_{\alpha}b_{\alpha}=\sum_{\pmb{k}}b^{\dagger}_{\pmb{k}}b_{\pmb{k}}$. For the anomalous hopping terms 
	\begin{equation}
		\begin{aligned}
			\sum_{\langle\alpha\beta\rangle}a_{\alpha}b_{\beta} &= \frac{2}{N}\sum_{\langle\alpha\beta\rangle}\sum_{\pmb{k}}e^{i\pmb{k}\cdot \pmb{r}_{\alpha}}a_{\pmb{k}}\sum_{\pmb{k}^{'}}e^{i\pmb{k}^{'}\cdot \pmb{r}_{\beta}}b_{\pmb{k}^{'}}\\
			&=\frac{2}{N}\sum_{\alpha,\pmb{\delta}}\sum_{\pmb{k}}e^{i\pmb{k}\cdot \pmb{r}_{\alpha}}a_{\pmb{k}}\sum_{\pmb{k}^{'}}e^{i\pmb{k}^{'}\cdot (\pmb{r}_{\alpha}+\pmb{\delta})}b_{\pmb{k}^{'}} \\
			&=\sum_{\pmb{\delta}}\sum_{\pmb{k}\pmb{k}^{'}}e^{i\pmb{k}^{'}\cdot \pmb{\delta}}a_{\pmb{k}}b_{\pmb{k}^{'}}\frac{2}{N}\sum_{\alpha}e^{i(\pmb{k}+\pmb{k}^{'})\cdot \pmb{r}_{\alpha}} \\
			&=\sum_{\pmb{k}}\gamma_{\pmb{k}}^{*}a_{\pmb{k}}b_{-\pmb{k}},
		\end{aligned}
	\end{equation}
	where $\gamma_{\pmb{k}}^{*}=(\sum_{\pmb{\delta}}e^{-i\pmb{k}\cdot \pmb{\delta}})$, and $\pmb{\delta}$ is the nearest-neighbor vector. Similarly, $\sum_{\langle \alpha \beta \rangle}a_{\alpha}^{\dagger}b_{\beta}^{\dagger}=\sum_{\pmb{k}}\gamma_{\pmb{k}}a_{\pmb{k}}^{\dagger}b_{-\pmb{k}}^{\dagger}$. The treatment for the next-nearest-neighbor normal terms is similar 
	\begin{equation}
		\begin{aligned}
			\sum_{\langle\langle\alpha\beta\rangle\rangle}-iJ_{\alpha\beta}^{(1)}S^{2} a_{\beta}^{\dagger} a_{\alpha}&=
			\sum_{\langle\langle\alpha\beta\rangle\rangle}-iJ_{\alpha\beta}^{(1)}S^{2}
			\sum_{\pmb{k}^{'}} e^{i\pmb{k}^{'}\cdot r_{\beta}} a_{\pmb{k}^{'}}^{\dagger}
			\sum_{\pmb{k}} e^{i\pmb{k}\cdot r_{\alpha}} a_{\pmb{k}} \\ 
			&=\sum_{\alpha,\pmb{d}} -iJ_{\alpha\beta}^{(1)}S^{2} \sum_{\pmb{k}^{'}} \sum_{\pmb{k}}a_{\pmb{k}^{'}}^{\dagger}a_{\pmb{k}}e^{-i\pmb{k}^{'}\cdot (r_{\alpha}+\pmb{d})}e^{i\pmb{k}\cdot r_{\alpha}}\\
			&=-\sum_{\pmb{k}}a_{\pmb{k}}^{\dagger}a_{\pmb{k}}\sum_{\pmb{d}}iJ_{\alpha\beta}^{(1)}S^{2} e^{i\pmb{k}\cdot\pmb{d}}.
		\end{aligned}
	\end{equation}
	Note that $J_{\alpha\beta}^{(1)}$ only depends on the next-nearest-neighbor vector $\pmb{d}$. We obtain 
	\begin{equation}
		\begin{aligned}
			\sum_{\langle\langle\alpha\beta\rangle\rangle}-iJ_{\alpha\beta}^{(1)}S^{2}(a_{\beta}^{\dagger}a_{\alpha}-a_{\alpha}^{\dagger}a_{\beta})&=\sum_{\pmb{k}}a_{\pmb{k}}^{\dagger}a_{\pmb{k}}\sum_{\pmb{d}}-iJ_{\alpha\beta}^{(1)}S^{2}(e^{-i\pmb{k}\cdot\pmb{d}}-e^{i\pmb{k}\cdot\pmb{d}}) \\
			&=-\sum_{\pmb{k}}D(\pmb{k})a_{\pmb{k}}^{\dagger}a_{\pmb{k}},
		\end{aligned}
	\end{equation}
	where $D(\pmb{k})=2S^{2}\sum_{\pmb{d}}J_{\alpha\beta}^{(1)}\sin(\pmb{k}\cdot\pmb{d})$.
	
	The combined $k$-space Hamiltonian is 
	\begin{equation}
		\mathcal{H}_{eff}=\sum_{\pmb{k}}\left[(zJS-D(\pmb{k}))a_{\pmb{k}}^{\dagger}a_{\pmb{k}}+(zJS+D(\pmb{k}))b_{\pmb{k}}^{\dagger}b_{\pmb{k}}+J_{0,\perp}S(\gamma_{\pmb{k}}a_{\pmb{k}}^{\dagger}b_{-\pmb{k}}^{\dagger}+\gamma_{\pmb{k}}^{*}a_{\pmb{k}}b_{-\pmb{k}})\right].
	\end{equation}
	
	To handle two-excitation terms like $a_{\pmb{k}}^{\dagger}b_{-\pmb{k}}^{\dagger}$, we introduce the 4-component Nambu spinor 
	\begin{equation}
		\Psi_{\pmb{k}}=\begin{pmatrix}a_{\pmb{k}}\\ b_{\pmb{k}}\\ a_{-\pmb{k}}^{\dagger}\\ b_{-\pmb{k}}^{\dagger}\end{pmatrix}, \quad \Psi_{\pmb{k}}^{\dagger}=\begin{pmatrix}a_{\pmb{k}}^{\dagger}&b_{\pmb{k}}^{\dagger}&a_{-\pmb{k}}&b_{-\pmb{k}}\end{pmatrix}.
	\end{equation}
	Constructing the matrix form  $\mathcal{H}_{eff}=\frac{1}{2}\sum_{\pmb{k}}\Psi_{\pmb{k}}^{\dagger}H(\pmb{k})\Psi_{\pmb{k}}$.
	
	Filling in the derived matrix elements, the complete Hamiltonian matrix is 
	\begin{equation}
		H_{4\times4}(\pmb{k}) = \begin{pmatrix} zJS-D(\pmb{k}) & 0 & 0 & J_{0,\perp}S\gamma_{\pmb{k}} \\ 0 & zJS+D(\pmb{k}) & J_{0,\perp}S\gamma_{\pmb{k}}^* & 0 \\ 0 & J_{0,\perp}S\gamma_{\pmb{k}} & zJS+D(\pmb{k}) & 0 \\ J_{0,\perp}S\gamma_{\pmb{k}}^* & 0 & 0 & zJS-D(\pmb{k}) \end{pmatrix}.
	\end{equation}
	
	However, there are two choices for Nambu basis vectors 
	\begin{equation}
		\Psi_{\pmb{k}}=\begin{pmatrix}a_{\pmb{k}}\\ b_{\pmb{k}}\\ a_{-\pmb{k}}^{\dagger}\\ b_{-\pmb{k}}^{\dagger}\end{pmatrix}, \quad \Psi_{\pmb{k}}^{\dagger}=\begin{pmatrix}a_{\pmb{k}}^{\dagger}&b_{\pmb{k}}^{\dagger}&a_{-\pmb{k}}&b_{-\pmb{k}}\end{pmatrix},
	\end{equation}
	\begin{equation}
		\Psi_{\pmb{k}}=\begin{pmatrix}a_{\pmb{k}}\\ b_{-\pmb{k}}^{\dagger}\end{pmatrix}, \quad \Psi_{\pmb{k}}^{\dagger}=\begin{pmatrix}a_{\pmb{k}}^{\dagger}&b_{-\pmb{k}}\end{pmatrix}.
	\end{equation}
	Constructing the matrix forms 
	
	Four-component 
	\begin{equation}
		H_{4\times4}(\pmb{k}) = \begin{pmatrix} zJS-D(\pmb{k}) & 0 & 0 & J_{0,\perp}S\gamma_{\pmb{k}} \\ 0 & zJS+D(\pmb{k}) & J_{0,\perp}S\gamma_{\pmb{k}}^* & 0 \\ 0 & J_{0,\perp}S\gamma_{\pmb{k}} & zJS+D(\pmb{k}) & 0 \\ J_{0,\perp}S\gamma_{\pmb{k}}^* & 0 & 0 & zJS-D(\pmb{k}) \end{pmatrix},
	\end{equation}
	
	Two-component 
	\begin{equation}
		H_{2\times2}(\pmb{k})  = \begin{pmatrix} zJS-D(\pmb{k}) & J_{0,\perp}S\gamma_{\pmb{k}} \\ J_{0,\perp}S\gamma_{\pmb{k}}^* & zJS-D(\pmb{k}) \end{pmatrix}.
	\end{equation}
	The pairing terms in $\mathcal{H}_{eff}$ are 
	\begin{equation}
		J_{0,\perp}S(\gamma_{\pmb{k}}a_{\pmb{k}}^{\dagger}b_{-\pmb{k}}^{\dagger}+\gamma_{\pmb{k}}^{*}a_{\pmb{k}}b_{-\pmb{k}}).
	\end{equation}
	These interactions only occur between $a_{\pmb{k}}$ and $b_{-\pmb{k}}$, as well as between $b_{\pmb{k}}$ and $a_{-\pmb{k}}$. The $H_{4\times4}(\pmb{k})$ matrix can thus be exactly decoupled into two independent $2\times2$ subspaces, with basis vectors $\Psi_{1,k} = (a_{\pmb{k}}, b_{-\pmb{k}}^\dagger)^T$ and $\Psi_{2,k} = (b_{\pmb{k}}, a_{-\pmb{k}}^\dagger)^T$. $H_{2\times2}(\pmb{k})$ is  the Hamiltonian matrix for subspace 1.
	
	Grouping the coupled operators together, we define a new arrangement order for the basis vectors 
	\begin{equation}
		\Psi'_{\pmb{k}} = \begin{pmatrix} a_{\pmb{k}} \\ b_{-\pmb{k}}^\dagger \\ b_{\pmb{k}} \\ a_{-\pmb{k}}^\dagger \end{pmatrix} = \begin{pmatrix} \Psi_{1,\pmb{k}} \\ \Psi_{2,\pmb{k}} \end{pmatrix}.
	\end{equation}
	Under this new basis $\Psi'_{\pmb{k}}$, the Hamiltonian becomes a block-diagonal matrix 
	\begin{equation}
		H'_{4\times4}(\pmb{k}) = \begin{pmatrix} H_{1}(\pmb{k}) & \mathbf{0} \\ \mathbf{0} & H_{2}(\pmb{k}) \end{pmatrix} = \begin{pmatrix} zJS-D(\pmb{k}) & J_{0,\perp}S\gamma_{\pmb{k}} & 0 & 0 \\ J_{0,\perp}S\gamma_{\pmb{k}}^* & zJS-D(\pmb{k}) & 0 & 0 \\ 0 & 0 & zJS+D(\pmb{k}) & J_{0,\perp}S\gamma_{\pmb{k}}^* \\ 0 & 0 & J_{0,\perp}S\gamma_{\pmb{k}} & zJS+D(\pmb{k}) \end{pmatrix}.
	\end{equation}
	
	For bosonic systems, the components of the Nambu spinor no longer satisfy the anticommutation relations of fermions, but rather obey bosonic commutation relations. Due to the restriction imposed by the commutation relation $[a, a^\dagger] = 1$, we cannot directly diagonalize the Hamiltonian matrix $H(\pmb{k})$, and must introduce the metric matrix $g$.
	For the four-component basis vector $\Psi_{\pmb{k}} = (a_{\pmb{k}}, b_{\pmb{k}}, a_{-\pmb{k}}^{\dagger}, b_{-\pmb{k}}^{\dagger})^T$, its commutation relations satisfy $[\Psi_{k, i}, \Psi_{k, j}^\dagger] = g_{ij}$. Through calculation, the metric matrix is 
	\begin{equation}
		g_{4\times4} = \text{diag}(1, 1, -1, -1) = \sigma_z \otimes \mathbb{I}_{2\times2}.
	\end{equation}
	Similarly, for the decoupled two-component basis vector $\Psi_{1,k} = (a_{\pmb{k}}, b_{-\pmb{k}}^\dagger)^T$, the metric matrix reduces to 
	\begin{equation}
		g_{2\times2} = \begin{pmatrix} 1 & 0 \\ 0 & -1 \end{pmatrix} = \sigma_z.
	\end{equation}
	
	For subspace 1, the corresponding matrix product is 
	\begin{equation}
		g_{2\times2} H_1(\pmb{k}) = \begin{pmatrix} 1 & 0 \\ 0 & -1 \end{pmatrix} \begin{pmatrix} A_{\pmb{k}} & V_{\pmb{k}} \\ V_{\pmb{k}}^* & A_{\pmb{k}} \end{pmatrix} = \begin{pmatrix} A_{\pmb{k}} & V_{\pmb{k}} \\ -V_{\pmb{k}}^* & -A_{\pmb{k}} \end{pmatrix},
	\end{equation}
	where $A_{\pmb{k}} = zJS - D(\pmb{k})$ and $V_{\pmb{k}} = J_{0,\perp}S\gamma_{\pmb{k}}$.
	Solving its characteristic equation yields the first branch of magnons 
	\begin{equation}
		\omega_1(\pmb{k}) = \sqrt{[zJS - D(\pmb{k})]^2 - |J_{0,\perp}S\gamma_{\pmb{k}}|^2}.
	\end{equation}
	Similarly, for subspace 2, its corresponding generalized eigenvalue equation gives 
	\begin{equation}
		\omega_2(\pmb{k}) = \sqrt{[zJS + D(\pmb{k})]^2 - |J_{0,\perp}S\gamma_{-\pmb{k}}|^2}.
	\end{equation}
	
	\section{Elliptically Polarized Light}

	For elliptically polarized light, we write the electric field as
	\begin{equation}
		\mathbf{E}(t) = (E_x \cos \Omega t, E_y \sin \Omega t, 0).
	\end{equation}
	Correspondingly, the cross product with the directional vector $\hat{z}$ becomes 
	\begin{equation}
		\mathbf{E}(t) \times \hat{z}  = (E_y \sin \Omega t, -E_x \cos \Omega t, 0).
	\end{equation}
	When calculating the AC phase from lattice site $\alpha$ to $\beta$, with bond directional angle $\phi_{\alpha\beta}$ and bond length $a$ 
	\begin{equation}
		\theta_{\alpha\beta}(t) = \frac{g\mu_B}{\hbar c^2} \int_{r_\alpha}^{r_\beta} \left( \mathbf{E}(t) \times \hat{z} \right) \cdot d\boldsymbol{\ell}.
	\end{equation}
	Taking the dot product with $\boldsymbol{\ell} = a(\cos \phi_{\alpha\beta}, \sin \phi_{\alpha\beta}, 0)$ yields 
	\begin{equation}
		\theta_{\alpha\beta}(t) = \frac{g\mu_B a}{\hbar c^2} (E_y \sin \Omega t \cos \phi_{\alpha\beta} - E_x \cos \Omega t \sin \phi_{\alpha\beta}).
	\end{equation}
	Using a trigonometric identity, this time-dependent phase is rewritten as 
	\begin{equation}
		\theta_{\alpha\beta}(t) = \lambda_{\alpha\beta} \sin(\Omega t - \tilde{\phi}_{\alpha\beta}).
	\end{equation}
	The dimensionless coupling strength is no longer a constant $\lambda$, but a variable dependent on the bond direction 
	\begin{equation}
		\lambda_{\alpha\beta} = \frac{g\mu_B a}{\hbar c^2} \sqrt{E_y^2 \cos^2 \phi_{\alpha\beta} + E_x^2 \sin^2 \phi_{\alpha\beta}},
	\end{equation}
	and the relative angle is no longer equal to the lattice geometric angle $\phi_{\alpha\beta}$, but satisfies 
	\begin{equation}
		\tan \tilde{\phi}_{\alpha\beta} = \frac{E_x}{E_y} \tan \phi_{\alpha\beta}.
	\end{equation}
	Under elliptically polarized light, upon substituting $\lambda_{\alpha\beta}$ into the zeroth-order Bessel function, bonds in different directions acquire different transverse exchange strengths 
	\begin{equation}
		J_{0,\perp}^{\alpha\beta} = J \mathcal{J}_0(\lambda_{\alpha\beta}).
	\end{equation}
	The zeroth-order effective Hamiltonian is modified to 
	\begin{equation}
		\mathcal{H}_{eff}^{(0)} = \sum_{\langle\alpha\beta\rangle} [J_{0,\perp}^{\alpha\beta} (S_\alpha^x S_\beta^x + S_\alpha^y S_\beta^y) + J S_\alpha^z S_\beta^z].
	\end{equation}
	In a honeycomb lattice such as graphene, the $C_3$ rotational symmetry of the system is broken because $\lambda_{\alpha\beta}$ along the three nearest-neighbor bonds $\delta_1, \delta_2, \delta_3$ are unequal.
	
	The first-order correction generates an effective DM interaction 
	\begin{equation}
		\mathcal{H}_{eff}^{(1)} = \sum_{\langle\langle\alpha\beta\rangle\rangle} J_{\alpha\beta}^{(1)} \mathbf{S}_{\gamma} \cdot (\mathbf{S}_{\alpha} \times \mathbf{S}_{\beta}).
	\end{equation}
	For elliptically polarized light, the calculation of the DMI strength $J_{\alpha\beta}^{(1)}$ 
	\begin{equation}
		J_{\alpha\beta}^{(1)} = \sum_{n=1}^\infty \frac{1}{n\Omega}2 J_{n,\perp}^2 \sin(n\Phi_{\alpha\beta}),
	\end{equation}
	undergoes two modifications:  the $J_{n,\perp}^2$ term in the coefficient becomes a product associated with the two path segments involved in the transition ($\alpha\rightarrow\gamma$ and $\gamma\rightarrow\beta$), i.e., $J \mathcal{J}_n(\lambda_{\alpha\gamma}) \cdot J \mathcal{J}_n(\lambda_{\gamma\beta})$. The phase difference $\Phi_{\alpha\beta}$ is replaced by the twisted angle  $\tilde{\Phi}_{\alpha\beta} = \tilde{\phi}_{\alpha\gamma} - \tilde{\phi}_{\beta\gamma}$.
	
	The modified Floquet DMI strength formula is 
	\begin{equation}
		J_{\alpha\beta}^{(1)} = 2J^2 \sum_{n=1}^\infty \frac{1}{n\Omega} \mathcal{J}_n(\lambda_{\alpha\gamma}) \mathcal{J}_n(\lambda_{\gamma\beta}) \sin(n\tilde{\Phi}_{\alpha\beta}).
	\end{equation}
	The DMI induced by circularly polarized light has uniform strength across all next-nearest-neighbor bonds. In contrast, the DMI induced by elliptically polarized light exhibits varying strengths across different next-nearest-neighbor bonds.
	
	\section{Square Lattice Model}
	For an isotropic square lattice, due to the existence of the $[C_{2\perp} ||C_4]$ operation, the condition $[C_{2\perp} ||C_4]^2=[E ||C_2]$ guarantees the even-parity characteristic of magnons $\omega_\uparrow(\pmb{k})=\omega_\uparrow(-k)$. Therefore, by varying the exchange coupling strengths $J_{ij}$ between nearest-neighbor lattice sites, we break $[C_{2\perp} ||C_4]$ while preserving $[C_2||P]$. The application of a light field further breaks $[C_{2\perp}\mathcal{T} || \mathcal{T}]$, thereby enabling the magnon spectrum to satisfy the odd-parity feature $\omega_\uparrow(\pmb{k})=\omega_\downarrow(-k)$.
	
\begin{figure}[htbp]
		\centering 
		\includegraphics[width=0.7\linewidth]{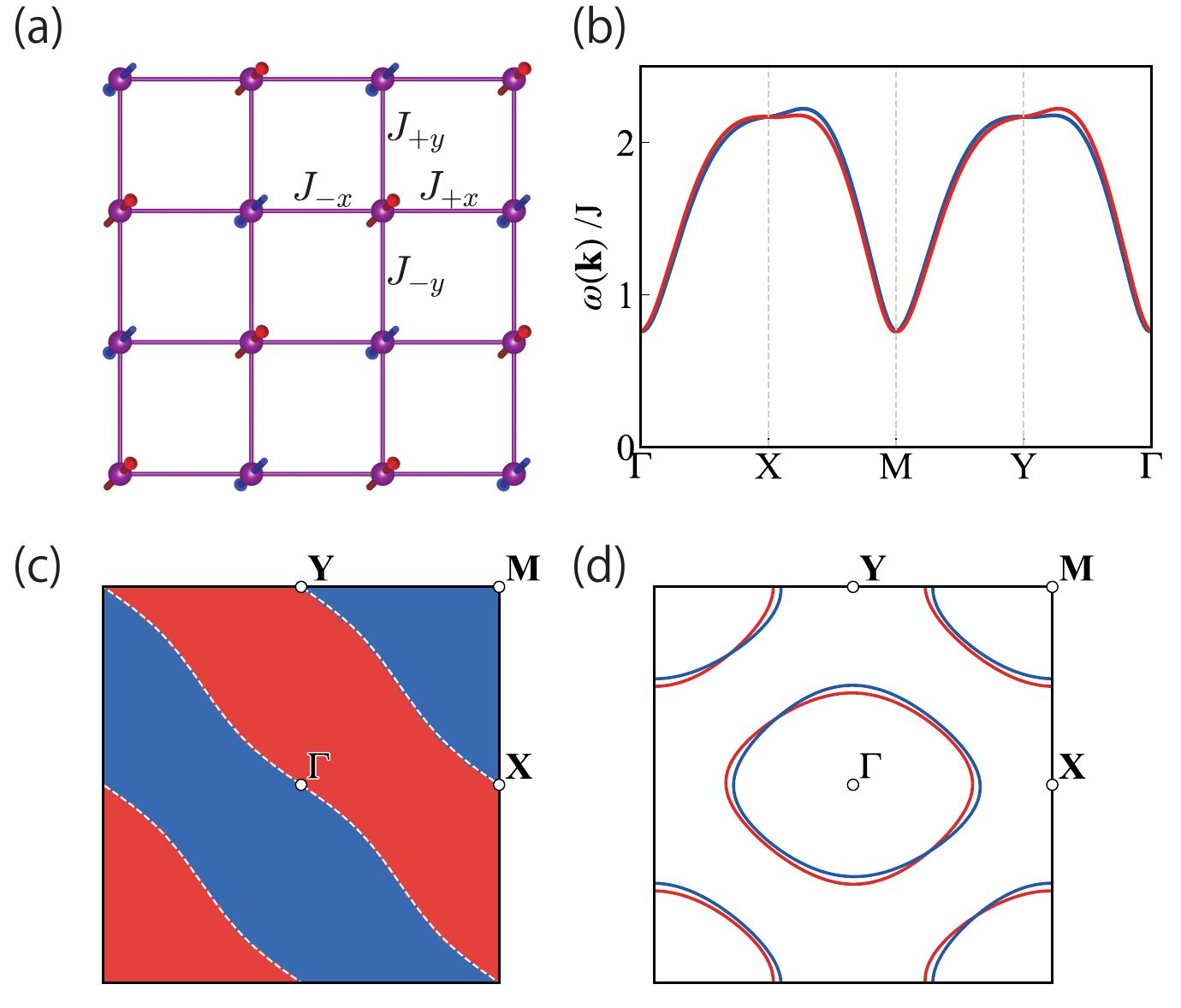}
		\caption{(a) Schematic of the square lattice and interaction parameters, where $S = 1/2$. The four antiferromagnetic exchange bonds with different strengths $J_{+x} = 0.8, J_{-x} = 1.0, J_{+y} = 1.2, J_{-y} = 1.4$ break $[C_2||C_4]$. The light-field driving parameter is $\lambda=0.5,\hbar\Omega/J=1$. (b) Band structure along high-symmetry paths. (c) Spin splitting distribution within the first Brillouin zone. (d) Momentum states in the Brillouin zone at the cut-plane energy $\omega(k)/J = 1.8$ of the dispersion curve, exhibiting a $p$-wave distribution.}
		\label{pic1}
	\end{figure}

	In an anisotropic system, the nearest-neighbor exchange interactions along the four directions of the square lattice can be entirely independent. Assuming a lattice constant $a$, we define four nearest-neighbor vectors in different directions 
	\begin{equation}
		\boldsymbol{\delta}_{+x} = (a, 0), \quad \boldsymbol{\delta}_{+y} = (0, a), \quad \boldsymbol{\delta}_{-x} = (-a, 0), \quad \boldsymbol{\delta}_{-y} = (0, -a).
	\end{equation}
	The corresponding nearest-neighbor antiferromagnetic coupling constants are denoted as $J_{+x}, J_{+y}, J_{-x}, J_{-y}$, as illustrated in the lattice schematic in Fig.~\ref{pic1}(a). The total nearest-neighbor coupling strength of the system is defined as $\Sigma_J = J_{+x}+ J_{+y}+ J_{-x}+ J_{-y}$.
	
	In the absence of a driving light field, the static Heisenberg Hamiltonian of the system is 
	\begin{equation}
		\mathcal{H}_0 = \sum_{\alpha \in A} \left( J_{+x} \mathbf{S}_\alpha \cdot \mathbf{S}_{\alpha+\boldsymbol{\delta}_1} + J_{+y} \mathbf{S}_\alpha \cdot \mathbf{S}_{\alpha+\boldsymbol{\delta}_2} + J_{-x} \mathbf{S}_\alpha \cdot \mathbf{S}_{\alpha+\boldsymbol{\delta}_3} + J_{-y} \mathbf{S}_\alpha \cdot \mathbf{S}_{\alpha+\boldsymbol{\delta}_4} \right).
	\end{equation}
	
	After performing the Holstein-Primakoff transformation and converting to momentum $\mathbf{k}$ space, the structure factor $\gamma_\mathbf{k}$ is asymmetric
	\begin{equation}
		\gamma_\mathbf{k} = J_{+x} e^{i k_x a} + J_{+y} e^{i k_y a} + J_{-x} e^{-i k_x a} + J_{-y} e^{-i k_y a}.
	\end{equation}
	Because $J_{+x} \neq J_{-x}$ and $J_{+y} \neq J_{-y}$, the spatial inversion symmetry $\mathcal{P}$ of the system is broken.
	
	Upon introducing the 4-component Nambu spinor $\Psi_{\mathbf{k}} = (a_\mathbf{k}, b_\mathbf{k}, a_{-\mathbf{k}}^\dagger, b_{-\mathbf{k}}^\dagger)^T$, the complete $H_{4\times4}(\mathbf{k})$ matrix can actually be exactly decoupled into two independent $2\times2$ subspaces. We select the basis vector for the first subspace as $\Psi_{1,\mathbf{k}} = (a_\mathbf{k}, b_{-\mathbf{k}}^\dagger)^T$. For the undriven system, the Hamiltonian matrix for subspace 1 is 
	\begin{equation}
		H_{2\times 2}^{(0)}(\mathbf{k}) = \begin{pmatrix} \Sigma_J S & J_{0,\perp}S \gamma_\mathbf{k} \\ J_{0,\perp}S \gamma_\mathbf{k}^* & \Sigma_J S \end{pmatrix}.
	\end{equation}
	
	When the system is exposed to a time-periodic circularly polarized light field with frequency $\Omega$ and intensity parameter $\lambda$, a Floquet topological phase transition occurs via the Aharonov-Casher effect. The light field induces an effective next-nearest-neighbor DMI interaction through the first-order term in the high-frequency expansion.
	
	Owing to the anisotropy of the nearest-neighbor interactions, the asymmetry of the transition paths is directly inherited by the DMI. We define four next-nearest-neighbor diagonal vectors $\mathbf{d}_1 = \boldsymbol{\delta}_{+x} - \boldsymbol{\delta}_{+y}$, $\mathbf{d}_2 = \boldsymbol{\delta}_{+y} - \boldsymbol{\delta}_{-x}$, $\mathbf{d}_3 = \boldsymbol{\delta}_{-x} - \boldsymbol{\delta}_{-y}$, $\mathbf{d}_4 = \boldsymbol{\delta}_{-y} - \boldsymbol{\delta}_{+x}$.
	In the anisotropic model, the mass term $D(\mathbf{k})$ in momentum space is no longer a simple superposition of sines; its expanded form relies on the specific products of adjacent bonds 
	\begin{equation}
		\begin{aligned}
			D(\mathbf{k}) =  \sum_{n=1}^{\infty}2S^2\mathcal{J}_n(\lambda)^2\sin(n\Phi_{\alpha\beta}) \frac{1}{n\Omega} \left[ \right.&J_{+x} J_{+y} \sin(\mathbf{k} \cdot \mathbf{d}_1) + J_{+y} J_{-x} \sin(\mathbf{k} \cdot \mathbf{d}_2) + \\ 
			&J_{-x} J_{-y} \sin(\mathbf{k} \cdot \mathbf{d}_3) + J_{-y} J_{+x} \sin(\mathbf{k} \cdot \mathbf{d}_4) \left.\right].
		\end{aligned}
	\end{equation}
	
	Integrating this back into the effective Hamiltonian of subspace 1 yields the final light-field-driven matrix 
	\begin{equation}
		H_{2\times 2}(\mathbf{k}) = \begin{pmatrix} \Sigma_JS - D(\mathbf{k}) & J_{0,\perp} S \gamma_\mathbf{k} \\ J_{0,\perp} S \gamma_\mathbf{k}^* & \Sigma_J S + D(\mathbf{k}) \end{pmatrix}.
	\end{equation}
	Similar to the hexagonal lattice, the effective Hamiltonian for subspace 2 can be obtained analogously. Calculating the magnon spectrum along the high-symmetry path $\Gamma-X-M-Y-\Gamma$ produces the odd-parity spin splitting as shown in Fig.~\ref{pic1}(b)-(d).

	\section{A-type AFM Magnon Model}
	
	\subsection{AA Stacking}
	\begin{figure}[htbp]
		\centering 
		\includegraphics[width=0.7\linewidth]{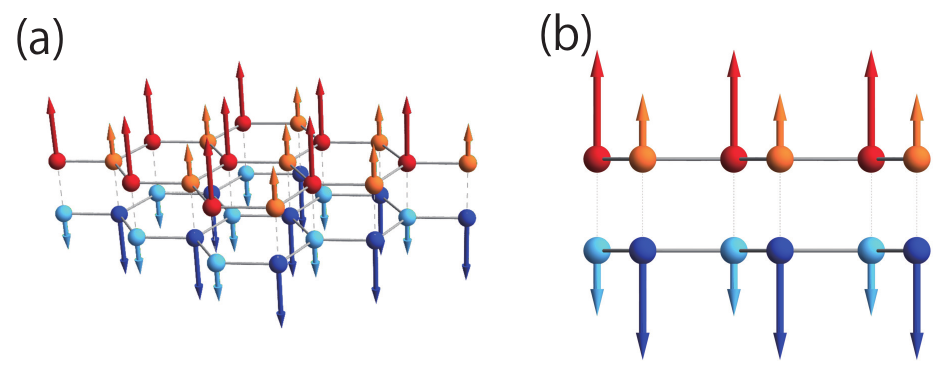}
		\caption{(a) Schematic of the A-type AFM $S_1\neq S_2$ lattice, showing the antiferromagnetic configuration in a bird's-eye view. The spins in the upper layer are ferromagnetically aligned within the layer, while the spins in the lower layer point in the opposite direction, forming an interlayer antiferromagnetic coupling. (b) Simplified side view.}
		\label{pic2}
	\end{figure}
	
	We extend the model to an AA-stacking bilayer honeycomb lattice, with its magnetic order set to A-type AFM. The intralayer coupling is ferromagnetic, lattice sites A and B carry unequal spin magnitudes $S_1 \neq S_2$, and the interlayer spins are antiparallel, as depicted in Fig.~\ref{pic2}.
	The unit cell contains 4 magnetic sublattices. The upper layer $L_1$ has spin up, $S_{A1}^z = +S_1, \quad S_{B1}^z = +S_2$. The lower layer $L_2$ has spin down, $S_{A2}^z = -S_2, \quad S_{B2}^z = -S_1$.
	Because the ground-state magnetization directions of the upper and lower layers are opposite, two sets of HP transformations with opposite orientations need to be applied. The HP transformations vary for spin-up and spin-down sites. For the spin-up $L_1$ layer 
	\begin{equation}
		S_\alpha^z = S_\alpha - a_\alpha^\dagger a_\alpha,$$$$S_\alpha^+ \approx \sqrt{2S_\alpha} a_\alpha, \quad S_\alpha^- \approx \sqrt{2S_\alpha} a_\alpha^\dagger,
	\end{equation}
	and for the spin-down $L_2$ layer 
	\begin{equation}
		S_\beta^z = -S_\beta + b_\beta^\dagger b_\beta,$$$$S_\beta^+ \approx \sqrt{2S_\beta} b_\beta^\dagger, \quad S_\beta^- \approx \sqrt{2S_\beta} b_\beta.
	\end{equation}
	
	The total Hamiltonian of the system is $H = H_{\text{intra}} + H_{\text{inter}}$. To solve for the bulk bands, theoretically, an 8-dimensional Nambu spinor needs to be introduced because the interlayer antiferromagnetic coupling generates anomalous pairing terms such as $a_{\pmb{k}}^\dagger b_{-\pmb{k}}^\dagger$ 
	\begin{equation}
		\Psi_{\pmb{k}} = (a_{A1,k}, a_{B1,k}, b_{A2,k}, b_{B2,k}, a_{A1,-k}^\dagger, a_{B1,-k}^\dagger, b_{A2,-k}^\dagger, b_{B2,-k}^\dagger)^T.
	\end{equation}
	In this case, the total Hamiltonian is written as $H = \frac{1}{2} \sum_{\pmb{k}} \Psi_{\pmb{k}}^\dagger H_{BdG}(\pmb{k}) \Psi_{\pmb{k}}$, where 
	\begin{equation}
		H_{BdG}(\pmb{k}) = \begin{pmatrix} M(\pmb{k}) & N(\pmb{k}) \\ N(\pmb{k})^\dagger & M(-k)^* \end{pmatrix},
	\end{equation}
	\begin{equation}
		H_{BdG}(\mathbf{k}) = \begin{pmatrix}
			H_{L1}(\mathbf{k}) & 0 & 0 & V_{inter} \\
			0 & H_{L2}(\mathbf{k}) & V_{inter}^\dagger & 0 \\
			0 & V_{inter} & H_{L1}^\dagger(-\mathbf{k}) & 0 \\
			V_{inter}^\dagger & 0 & 0 & H_{L2}^\dagger(-\mathbf{k})
		\end{pmatrix}.
	\end{equation}
	The effective Hamiltonian matrix for the spin-up subspace is 
	\begin{equation}
		H_{\uparrow}^{(4 \times 4)} = \begin{pmatrix}
			H_{L1}(\mathbf{k}) & V_{inter} \\
			V_{inter}^\dagger & H_{L2}^\dagger(-\mathbf{k})
		\end{pmatrix},
	\end{equation}
	which explicitly expands to 
	\begin{equation}
		H_{\uparrow}^{(4 \times 4)} = \begin{pmatrix}
			\mathcal{E}_{A1} & J_{intra}\gamma_\mathbf{k} & 0 & \Delta_{inter} \\
			J_{intra}\gamma_\mathbf{k}^\dagger & \mathcal{E}_{B1} & \Delta_{inter}^\dagger & 0 \\
			0 & \Delta_{inter} & \mathcal{E}_{A2} & J_{intra}\gamma_{-\mathbf{k}}^\dagger \\
			\Delta_{inter}^\dagger & 0 & J_{intra}\gamma_{-\mathbf{k}} & \mathcal{E}_{B2}
		\end{pmatrix}.
	\end{equation}
	Here, $\text{intra}$ denotes intralayer hopping, and $\text{inter}$ denotes interlayer hopping.
	In the limit of weak interlayer antiferromagnetic coupling, we focus on the hopping matrix $M(\pmb{k})$ containing only normal pairing terms, which governs the topological phase transition. Since there is no particle-number-conserving hopping that creates and annihilates magnon pairs between layers, and only anomalous pairing exists, $M(\pmb{k})$ is strictly block-diagonalized as 
	\begin{equation}
		M(\pmb{k}) = \begin{pmatrix} H_{L1}(\pmb{k}) & 0 \\ 0 & H_{L2}(\pmb{k}) \end{pmatrix}.
	\end{equation}
	Driven by circularly polarized light, the light field introduces next-nearest-neighbor DMI into the Hamiltonians $H_{L1}(\pmb{k})$ and $H_{L2}(\pmb{k})$ of each layer due to the broken time-reversal symmetry.
	
	The intralayer interaction is ferromagnetic 
	\begin{equation}
		H_{\text{intra}} = -J_1 \sum_{\langle \alpha,\beta \rangle} \pmb{S}_\alpha \cdot \pmb{S}_\beta.
	\end{equation}
	Expanding the dot product $\pmb{S}_\alpha \cdot \pmb{S}_\beta = S_\alpha^z S_\beta^z + \frac{1}{2}(S_\alpha^+ S_\beta^- + S_\alpha^- S_\beta^+)$. The latter two terms represent the hopping part 
	\begin{equation}
		\begin{aligned}
			H &= -\frac{J_1}{2} \sum_{\langle \alpha,\beta \rangle} (\sqrt{2S_1} a_{\alpha} \sqrt{2S_2} a_{\beta}^\dagger + \sqrt{2S_1} a_{\alpha}^\dagger \sqrt{2S_2} a_{\beta})\\
			&= -J_1 \sqrt{S_1 S_2} \sum_{\langle \alpha,\beta \rangle} (a_{\alpha}^\dagger a_{\beta} + h.c.).
		\end{aligned}
	\end{equation}
	Substituting the Fourier transform  $\sum_{\langle \alpha,\beta \rangle} a_{A1,\alpha}^\dagger a_{B1,\beta} = \sum_{\pmb{k}} a_{A1,k}^\dagger a_{B1,k} \sum_{\pmb{\delta}} e^{i \pmb{k}\cdot(\pmb{r}_\beta - \pmb{r}_\alpha)}$, we define the vector $\pmb{\delta}$ as the nearest-neighbor vector pointing from A to B.
	
	Intralayer nearest-neighbor ferromagnetic coupling between $A1$ ($S_1$, up) and 3 $B1$ atoms ($S_2$, up) 
	\begin{equation}
		-J_1 S_{A1}^z S_{B1}^z = -J_1 (S_1 - a_{A1}^\dagger a_{A1})(S_2 - a_{B1}^\dagger a_{B1}).
	\end{equation}
	After expansion, the term contributing to $a_{A1}^\dagger a_{A1}$ is $+J_1 S_2 a_{A1}^\dagger a_{A1}$. Since there are 3 nearest neighbors, the total contribution is $3 J_1 S_2$ 
	\begin{equation}
		\mathcal{E}_{A1}(\pmb{k})=3 J_1 S_2.
	\end{equation}
	Similarly, for $B1$ coupled to 3 $A1$ atoms 
	\begin{equation}
		\mathcal{E}_{B1}(\pmb{k})=3 J_1 S_1.
	\end{equation}
	The $2\times 2$ Hamiltonian for the single-layer ferromagnetic coupling is written as 
	\begin{equation}
		H_{L1}(\pmb{k}) = \begin{pmatrix} 3 J_1 S_2 & -J_1 \sqrt{S_1 S_2} \gamma_{\pmb{k}} \\ -J_1 \sqrt{S_1 S_2} \gamma_{\pmb{k}}^* & 3 J_1 S_1 \end{pmatrix}.
	\end{equation}
	Likewise, for the $L_2$ layer ferromagnetism 
	\begin{equation}
		H_{L2}(\pmb{k}) = \begin{pmatrix} 3 J_1 S_1 & -J_1 \sqrt{S_1 S_2} \gamma_{\pmb{k}} \\ -J_1 \sqrt{S_1 S_2} \gamma_{\pmb{k}}^* & 3 J_1 S_2 \end{pmatrix}.
	\end{equation}
	Due to the AA stacking, $\gamma_{\pmb{k}}$ is identical for both layers. 

	Taking $H_{L1}(\pmb{k})$ as an example, a mass term exists at the $K$ and $K'$ valleys in momentum space 
	\begin{equation}
		m= \frac{3J_1(S_1 - S_2)}{2}.
	\end{equation}
	Without an applied light field, the mass naturally opens the Dirac cones of the monolayer ferromagnet, leaving the system in a trivial insulating phase.
	
	Upon applying circularly polarized light, the system acquires next-nearest-neighbor DMI. The strength of the DMI depends on the three spins along the transition path. For an A-A transition involving site spin $S_1$, the intermediate site is a B spin $S_2$. Since the transition term includes $\sqrt{S_1}\sqrt{S_1} = S_1$, multiplying by the intermediate site's $S_B^z \approx S_2$ yields a total coefficient proportional to $S_1 S_2$. Similarly, for a B-B transition, the coefficient is proportional to $S_B S_B S_A^z = S_2 S_1$. This demonstrates that despite the unequal A and B spins, the Floquet DMI strengths they acquire remain symmetric. Introducing the DMI structure factor $D(\pmb{k})$ in momentum space, the complete effective Hamiltonian is 
	\begin{equation}
		H_{L1}(\pmb{k}) = \begin{pmatrix} 3 J_1 S_2 +D(\pmb{k})& -J_1 \sqrt{S_1 S_2} \gamma_{\pmb{k}} \\- J_1 \sqrt{S_1 S_2} \gamma_{\pmb{k}}^* & 3 J_1 S_1- D(\pmb{k}) \end{pmatrix},
	\end{equation}
	where $D(\mathbf{k}) = 2S_1S_2 \sum_{\mathbf{d}} J_{\alpha\beta}^{(1)} \sin(\mathbf{k} \cdot \mathbf{d})$ and $J_{\alpha\beta}^{(1)} = \sum_{n=1}^\infty \frac{1}{n\omega}2 J_{n,\perp}^2 \sin(n\Phi_{\alpha\beta})$.
	Considering the effect of the light field, it introduces next-nearest-neighbor DMI interactions in each ferromagnetic layer, with opposite DMI strengths at the K and K' points.
	Similarly, for the $L_2$ layer ferromagnetism 
	\begin{equation}
		H_{L2}(\pmb{k}) = \begin{pmatrix} 3 J_1 S_1  + D(\pmb{k})&- J_1 \sqrt{S_1 S_2} \gamma_{\pmb{k}} \\ -J_1 \sqrt{S_1 S_2} \gamma_{\pmb{k}}^* & 3 J_1 S_2 - D(\pmb{k})\end{pmatrix}.
	\end{equation}
	
	The effective Hamiltonian matrix for the spin-up subspace is 
	\begin{equation}
	H_{\uparrow}^{(4 \times 4)} = \begin{pmatrix}
		H_{L1}(\mathbf{k}) & V_{inter} \\
		V_{inter}^\dagger & H_{L2}^\dagger(-\mathbf{k})
	\end{pmatrix}.
	\end{equation}
	The off-diagonal block $V_{inter}$ represents the interlayer anomalous pairing  resulting from the A-type AFM coupling. Due to the AA stacking, the non-diagonal elements in $V_{inter}$ are independent of the momentum $\mathbf{k}$. This implies that the Floquet DMI originates exclusively from the intralayer hopping, and the interlayer coupling does not acquire photo-induced topological contributions.
	
	The interlayer anomalous pairing $V_{inter}$ merely introduces momentum-independent constant terms into the zeroth-order Hamiltonian and does not contribute to the topological phase transition of the system. Furthermore, retaining the interlayer coupling complicates the analytical derivation of the magnon energy eigenvalues without offering additional topological insights. The full bosonic BdG spectrum, including the interlayer coupling, is adiabatically connected to the reduced model without additional gap closing. Therefore, to transparently capture the core topological physics at an analytical level, the subsequent analysis is conducted under the weak interlayer coupling limit approximation
	\begin{equation}
		H_{\uparrow}^{(4 \times 4)} \approx \begin{pmatrix}
			H_{L1}(\mathbf{k}) & 0 \\
			0 & H_{L2}^\dagger(-\mathbf{k})
		\end{pmatrix}.
	\end{equation}
	\begin{figure}[htbp]
		\centering 
		\includegraphics[width=0.7\linewidth]{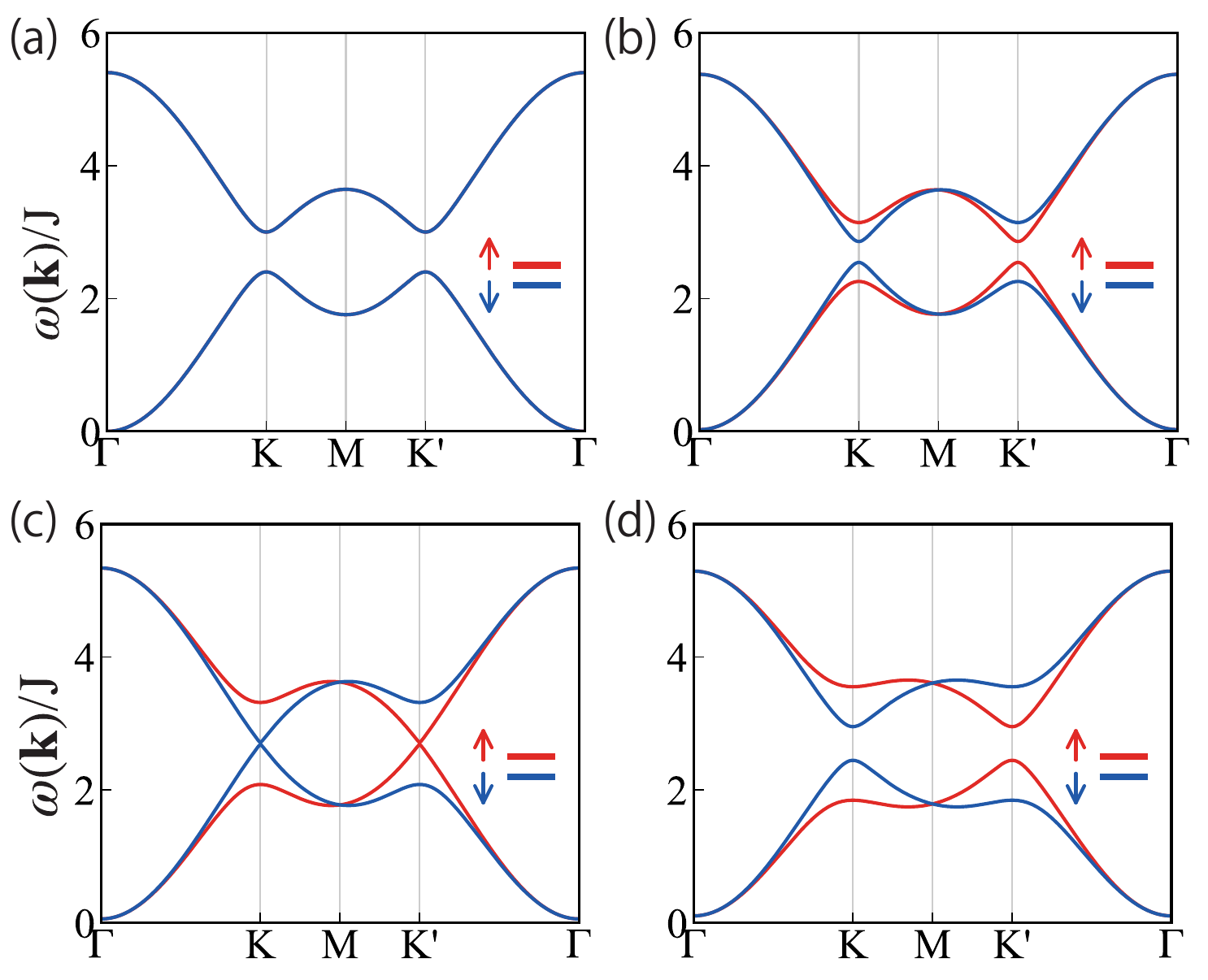}
		\caption{Magnon dispersion relations $\omega(\mathbf{k})/J$ of the AA stacking model along the high-symmetry path ($\Gamma-K-M-K'-\Gamma$) in the Brillouin zone under different light-field driving parameters $\lambda$. (a) No light-field driving $\lambda=0$, (b) $\lambda=0.2$, (c) $\lambda=0.3$, (d) $\lambda=0.4$. The red and blue solid lines represent the spin-up ($\uparrow$) and spin-down ($\downarrow$) magnon bands, respectively. The system parameters utilized in the calculation are $S_1=1$, $S_2=0.8$, and $\hbar\Omega/J =1$.}
		\label{pic3}
	\end{figure}
	\begin{figure}[htbp]
		\centering 
		\includegraphics[width=0.7\linewidth]{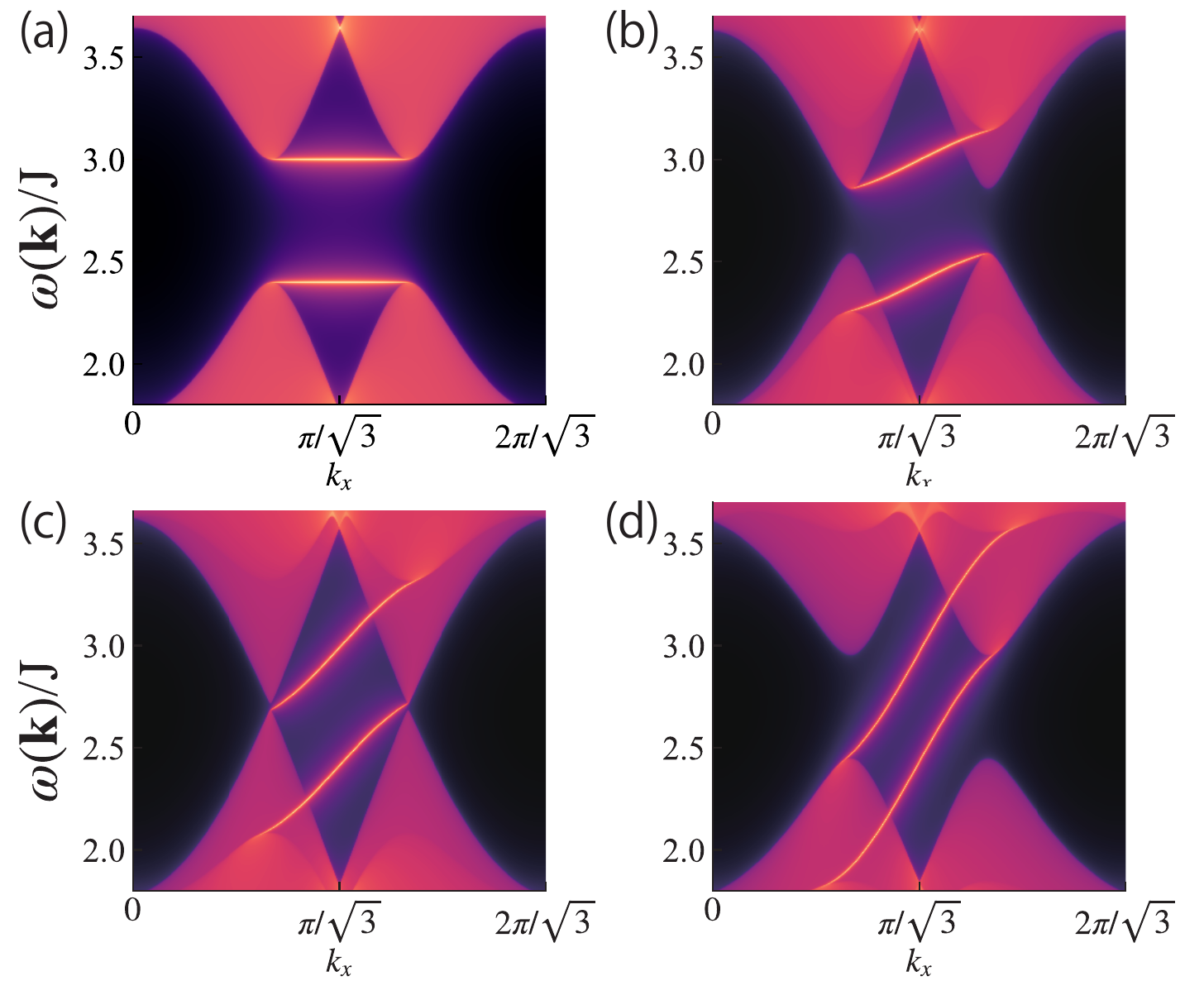}
		\caption{Magnon edge state spectra of the AA stacking model under different light-field driving parameters. (a) No light field ($\lambda=0$), (b) $\lambda=0.2$, (c) $\lambda=0.3$, (d) $\lambda=0.4$. The continuous diffuse regions represent bulk bands, whereas the bright, distinct lines traversing the bulk band gap correspond to topologically protected edge states. The system parameters used in the calculation are $S_1=1$, $S_2=0.8$, and $\hbar\Omega/J =1$.}
		\label{pic4}
	\end{figure}
	
	To intuitively demonstrate the topological phase transition process in the bilayer antiferromagnetic system, we calculated the bulk state dispersion and projected band structures under varying dimensionless light-field driving strengths $\lambda$, as shown in Fig.~\ref{pic3}. Under static conditions without illumination (Fig.~\ref{pic3}(a)), owing to the trivial mass introduced by the sublattice asymmetry $S_1 \neq S_2$, magnons exhibit a discernible bulk band gap at the $K$ and $K'$ valleys. The corresponding projected band structure also displays a clean band gap without any topological edge states. With the introduction of light-field driving, the time-reversal symmetry is broken, spin degeneracy is lifted, and odd-parity splitting emerges (Fig.~\ref{pic3}(b)). When the driving strength reaches the critical value $\lambda \approx 0.3$, the bulk band gap completely closes at the valleys, forming a Dirac cone , which signifies the system reaching a topological quantum critical point. Beyond this critical point ($\lambda = 0.4$), the topological mass dominates, causing a band inversion and the reopening of the bulk band gap. In the calculated projected band structure (Fig.~\ref{pic4}(d)), we clearly observe bright chiral edge states traversing the reopened bulk band gap, confirming that the system has successfully transitioned into a topological magnon phase characterized by a Chern number $C=2$.

	Expanding the structure factor at the $K / K'$ valleys of the Brillouin zone gives  $\gamma_{\mathbf{K}_\eta + \mathbf{q}} \approx \frac{3a}{2}(\eta q_x - i q_y)$, and $D(\mathbf{K}_\eta + \mathbf{q}) \approx \eta S_1S_2\sum_{\pmb{d}}J_{\alpha\beta}^{(1)} 3 \sqrt{3}$.
	Substituting these into the decomposition results yields the respective low-energy two-dimensional effective Dirac models for the two layers 
	\begin{equation}
		H_{L1}^\eta(\mathbf{q}) = v_F (\eta q_x \sigma_x + q_y \sigma_y) + m_0 \sigma_0 + M_1^\eta \sigma_z,
	\end{equation}
	\begin{equation}
		H_{L2}^\eta(\mathbf{q}) = v_F (\eta q_x \sigma_x + q_y \sigma_y) + m_0 \sigma_0 + M_2^\eta \sigma_z,
	\end{equation}
	where $\eta$ is the valley index, and the constant shift term is $m_0 = \frac{3J_1(S_1 + S_2)}{2}$. The mass terms are, respectively 
	\begin{equation}
		M_1^\eta = -\frac{3J_1(S_1 - S_2)}{2} + \eta S_1S_2\sum_{\pmb{d}}J_{\alpha\beta}^{(1)} 3 \sqrt{3},
	\end{equation}
	\begin{equation}
		M_2^\eta = \frac{3J_1(S_1 - S_2)}{2} + \eta S_1S_2\sum_{\pmb{d}}J_{\alpha\beta}^{(1)} 3 \sqrt{3}.
	\end{equation}
	
	For the two-dimensional effective Dirac model, the total Chern number of the $i$-th layer is determined by the sum of the local contributions from the two valleys 
	\begin{equation}
		C_{Li} = \frac{1}{2} \sum_{\eta=\pm} \eta \, \text{sgn}(M_i^\eta).
	\end{equation}
	
	Under weak light-field intensities, the sign of the effective mass is dominated by the trivial intrinsic mass, i.e., $\left| \frac{3J_1(S_1 - S_2)}{2} \right| > \left| S_1S_2\sum_{\pmb{d}}J_{\alpha\beta}^{(1)} 3 \sqrt{3} \right|$. For the $L_1$ layer, the mass terms at both valleys share the same sign, i.e., $\text{sgn}(M_1^+) = \text{sgn}(M_1^-) = -1$. Substituting this into the formula gives $C_{L1} = \frac{1}{2}[(+1)(-1) + (-1)(-1)] = 0$. Similarly, the total Chern number for the $L_2$ layer is $C_{L2} = 0$. The system is in a trivial insulating phase.
	
	Upon entering the strong-driving regime, the Floquet-induced topological mass becomes dominant. Specifically, for the $L_1$ layer, the sign of the mass term at the $K$ valley ($\eta=+1$) flips, making $\text{sgn}(M_1^+) = +1$, while for the $K'$ valley $\text{sgn}(M_1^-) = -1$. The single-valley Chern number contributions become $C_1^K = 1/2$ and $C_1^{K'} = 1/2$, and the total Chern number of the system abruptly transitions to $C_{L1} = \frac{1}{2}[(+1)(+1) + (-1)(-1)] = 1$.
	
	According to the specific formulation of the low-energy expansion, the intrinsic masses of the upper and lower layers carry opposite signs. Consequently, for the $L_2$ layer, band inversion transpires at the $K'$ valley ($\eta=-1$), causing a sign flip in $\text{sgn}(M_2^-)$. With $\text{sgn}(M_2^+) = -1$ and $\text{sgn}(M_2^-) = +1$, calculations similarly yield its total Chern number as $C_{L2} = 1$.
	
	Ultimately, the total Chern number for this bilayer system is $C_{total} = C_{L1} + C_{L2} = 2$.

	\subsection{AB Stacking}
	
	We consider an AB stacking bilayer honeycomb lattice featuring an A-type antiferromagnetic order. The intralayer coupling is ferromagnetic, and the interlayer coupling is antiferromagnetic. The upper and lower spin quantum numbers of the system are equal, $S_1 = S_2 = S$. The unit cell encompasses 4 magnetic sublattices. Under AB stacking, the $B_1$ atom of the upper $L_1$ layer is situated directly above the $A_2$ atom of the lower $L_2$ layer, whereas $A_1$ and $B_2$ are located at the centers of the hexagons and lack direct interlayer nearest neighbors. The ground-state magnetization configuration is defined as spin-up for the $L_1$ layer ($S_{A1}^z = +S, \quad S_{B1}^z = +S$) and spin-down for the $L_2$ layer ($S_{A2}^z = -S, \quad S_{B2}^z = -S$).
	
	The real-space spin Hamiltonian of the system comprises intralayer nearest-neighbor exchange, interlayer nearest-neighbor exchange, and the light-field-induced next-nearest-neighbor DMI 
	\begin{equation}
		H = H_{\text{intra}} + H_{\text{inter}} + H_{\text{Floquet}},
	\end{equation}
	\begin{equation}
		H_{\text{intra}} = -J_1 \sum_{\langle \alpha,\beta \rangle \in L_1} \mathbf{S}_\alpha \cdot \mathbf{S}_\beta -J_1 \sum_{\langle \alpha,\beta \rangle \in L_2} \mathbf{S}_\alpha \cdot \mathbf{S}_\beta,
	\end{equation}
	\begin{equation}
		H_{\text{inter}} = J_{\text{inter}} \sum_{\alpha \in B_1, \beta \in A_2} \mathbf{S}_\alpha \cdot \mathbf{S}_\beta.
	\end{equation}
	Substituting the HP transformation into $H_{\text{intra}}$ and $H_{\text{inter}}$ and retaining up to quadratic terms. The intralayer interaction contributes an equivalent potential field of $3J_1 S$ for each sublattice. Meanwhile, expanding the interlayer interaction $S_{B1}^z S_{A2}^z$ gives 
	\begin{equation}
		J_{\text{inter}} S_{B1}^z S_{A2}^z = J_{\text{inter}}(S - a_{B1}^\dagger a_{B1})(-S + b_{A2}^\dagger b_{A2}) \approx -J_{\text{inter}} S^2 + J_{\text{inter}} S (a_{B1}^\dagger a_{B1} + b_{A2}^\dagger b_{A2}).
	\end{equation}
	Due to the geometric asymmetry of the AB stacking, the interlayer potential field exclusively acts upon $B_1$ and $A_2$. The diagonal on-site energies of each sublattice are determined as 
	\begin{equation}
		\mathcal{E}_{A1} = 3J_1 S,$$$$\mathcal{E}_{B1} = 3J_1 S + J_{\text{inter}} S,$$$$\mathcal{E}_{A2} = 3J_1 S + J_{\text{inter}} S,$$$$\mathcal{E}_{B2} = 3J_1 S.
	\end{equation}
	The intralayer nearest-neighbor ferromagnetic hopping contributes conventional bosonic hopping terms $-J_1 S (a_i^\dagger a_j + \text{h.c.})$. The interlayer antiferromagnetic coupling $S_{B1}^x S_{A2}^x + S_{B1}^y S_{A2}^y = \frac{1}{2}(S_{B1}^+ S_{A2}^- + S_{B1}^- S_{A2}^+)$, upon expansion, produces anomalous pairing terms that violate particle number conservation 
	\begin{equation}
		\frac{J_{\text{inter}}}{2} (\sqrt{2S} a_{B1} \sqrt{2S} b_{A2} + \sqrt{2S} a_{B1}^\dagger \sqrt{2S} b_{A2}^\dagger) = J_{\text{inter}} S (a_{B1} b_{A2} + a_{B1}^\dagger b_{A2}^\dagger).
	\end{equation}
	Letting $\Delta = J_{\text{inter}} S$.
	Defining the in-plane next-nearest-neighbor lattice vectors as $\mathbf{d}_1, \mathbf{d}_2, \mathbf{d}_3$. After Fourier transformation, the contribution of the DMI term to the diagonal elements manifests as an odd-function structure factor $D(\mathbf{k})$ 
	\begin{equation}
		D(\mathbf{k}) = 2 S^2J_{\alpha\beta}^{(1)}\sum_{m=1}^3 \sin(\mathbf{k} \cdot \mathbf{d}_m).
	\end{equation}
	In collinear antiferromagnetic systems, the full 8×8 BdG matrix can be precisely decoupled into two independent $4 \times 4$ subspaces. We extract the Nambu spinor that characterizes the spin-up excitations 
	\begin{equation}
		\Phi_{\mathbf{k}} = (a_{A1,\mathbf{k}}, a_{B1,\mathbf{k}}, b_{A2,-\mathbf{k}}^\dagger, b_{B2,-\mathbf{k}}^\dagger)^T.
	\end{equation}
	The rigorous global $4 \times 4$ matrix $\mathcal{H}_{\mathbf{k}}$ should be articulated as 
	\begin{equation}
		\mathcal{H}_{\mathbf{k}} = \begin{pmatrix} \mathcal{E}_{A1} + D(\mathbf{k}) & -J_1 S \gamma_{\mathbf{k}} & 0 & 0 \\ -J_1 S \gamma_{\mathbf{k}}^* & \mathcal{E}_{B1} - D(\mathbf{k}) & \Delta & 0 \\ 0 & \Delta & \mathcal{E}_{A2} + D(\mathbf{k}) & -J_1 S \gamma_{-\mathbf{k}}^* \\ 0 & 0 & -J_1 S \gamma_{-\mathbf{k}} & \mathcal{E}_{B2} - D(\mathbf{k}) \end{pmatrix}.
	\end{equation}
	To ascertain the antiferromagnetic magnons, one must solve the generalized eigenvalue problem of $\Sigma_z \mathcal{H}_{\mathbf{k}}$, where $\Sigma_z = \text{diag}(1, 1, -1, -1)$ enforces the requisite bosonic commutation relations 
	\begin{equation}
		\Sigma_z \mathcal{H}_{\mathbf{k}} = \begin{pmatrix} \mathcal{E}_{A1} + D(\pmb{k}) & -J_1 S \gamma_{\mathbf{k}} & 0 & 0 \\ -J_1 S \gamma_{\mathbf{k}}^* & \mathcal{E}_{B1} - D(\pmb{k}) & \Delta & 0 \\ 0 & -\Delta & -(\mathcal{E}_{A2} + D(\pmb{k})) & J_1 S \gamma_{-\mathbf{k}}^* \\ 0 & 0 & J_1 S \gamma_{-\mathbf{k}} & -(\mathcal{E}_{B2} - D(\pmb{k})) \end{pmatrix}.
	\end{equation}
	
	To determine the critical light field for the topological phase transition, we examine the gap-closing behavior at the $K / K'$ valleys in momentum space.
	At the Dirac points, the structure factor $\gamma_{\mathbf{K}} = 0$. Here, the complex $4 \times 4$ matrix becomes block-diagonal. 
	The $A_1$ and $B_2$ sublattices do not partake in interlayer coupling; their energy eigenvalues are simply the diagonal elements themselves 
	\begin{equation}
		\begin{aligned}
			\omega_{A1} = 3J_1 S + D(\pmb{k}),
			\omega_{B2} = 3J_1 S - D(\pmb{k}).
		\end{aligned}
	\end{equation}
	The $B_1$ and $A_2$ sublattices are coupled via $\Delta$, forming an independent $2 \times 2$ subspace 
	\begin{equation}
		D_{\mathbf{K}}^{(B1,A2)} = \begin{pmatrix} 3J_1 S + \Delta - D(\pmb{k}) & \Delta \\ -\Delta & -(3J_1 S + \Delta + D(\pmb{k})) \end{pmatrix}.
	\end{equation}
	Solving for the analytical energies 
	\begin{equation}
		\omega_{\pm}^{(B1,A2)} = -D(\pmb{k}) \pm \sqrt{(3J_1 S + \Delta)^2 - \Delta^2}.
	\end{equation}
	Substituting back $\Delta = J_{\text{inter}} S$, the positive energy branch is 
	\begin{equation}
		\omega_{+}^{(B1,A2)} = -D(\pmb{k}) + \sqrt{9J_1^2 S^2 + 6J_1 J_{\text{inter}} S^2}.
	\end{equation}
	In the Dirac cone forged by the $L_1$ layer, the energy gap $\Delta_{\text{gap}}$ is dictated by the difference between $\omega_{A1}$ and $\omega_{+}^{(B1,A2)}$. The topological phase transition transpires at the band closing point, namely 
	\begin{equation}
		3J_1 S + 2D(\pmb{k}) = \sqrt{9J_1^2 S^2 + 6J_1 J_{\text{inter}} S^2},
	\end{equation}
	which yields the critical Floquet mass term incorporating the complete interlayer interaction 
	\begin{equation}
		D(\pmb{k}) = \frac{1}{2}\sqrt{9J_1^2 S^2 + 6J_1 J_{\text{inter}} S^2} - \frac{3}{2}J_1 S.
	\end{equation}
	
	\section{Material Calculations}
	All first-principles calculations in this paper were performed using the Vienna Ab initio Simulation Package (VASP)~\cite{kresseEfficientIterativeSchemes1996}. In the calculations, the exchange-correlation energy was treated using the Perdew-Burke-Ernzerhof (PBE) functional within the generalized gradient approximation (GGA)~\cite{perdewGeneralizedGradientApproximation1996}. To accurately describe the strong correlation effects of the localized $3d$ electrons of the transition metals in the system, we employed the DFT+$U$ method proposed by Dudarev et al.~\cite{dudarevElectronenergylossSpectraStructural1998}. Furthermore, since all the materials investigated in this paper exhibit prominent van der Waals layered characteristics, we incorporated the DFT-D3 method in all structural optimizations and static calculations to account for long-range van der Waals interactions~\cite{grimmeConsistentAccurateInitio2010}.
	
	To construct the effective spin Hamiltonian intended for describing low-energy magnon excitations, it is imperative to extract the magnetic exchange interaction parameters between lattice sites. We first employed the Wannier90 package to construct maximally localized Wannier functions based on the first-principles Bloch wave functions~\cite{mostofiWannier90ToolObtaining2008}, thereby accurately extracting the tight-binding Hamiltonian of the system. Subsequently, the open-source package \texttt{TB2J} was utilized to calculate and extract the exchange coupling parameters of the system from this tight-binding model~\cite{heTB2JPythonPackage2021}.
	
\begin{figure}[htbp]
		\centering 
		\includegraphics[width=0.7\linewidth]{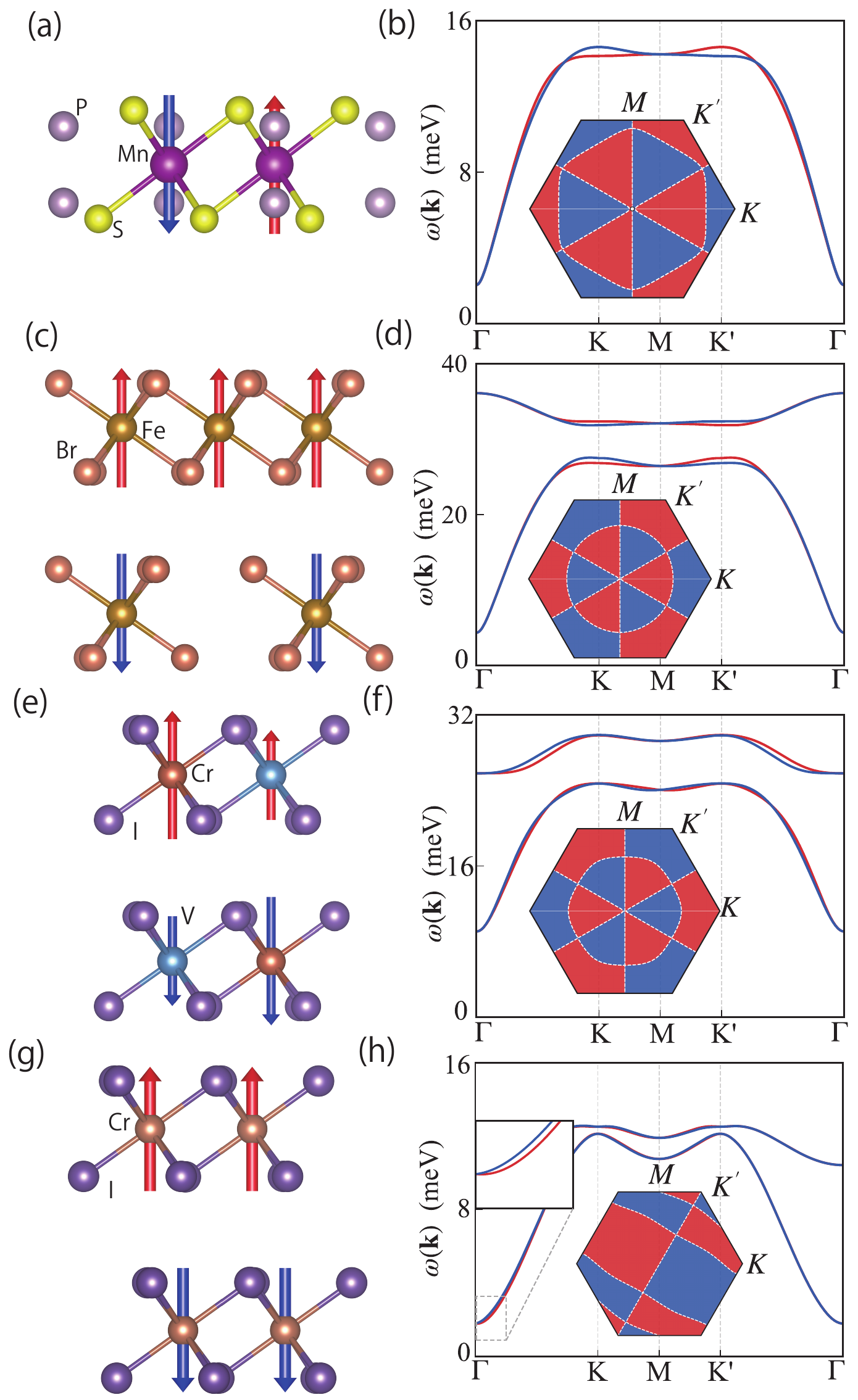}
		\caption{Schematics of the crystal structures and spin arrangements for (a) monolayer $\text{MnPS}_3$, (c) bilayer $\text{FeBr}_3$, (e) bilayer $\text{CrVI}_6$, and (g) slid bilayer $\text{CrI}_3$ (AB' stacking). The arrows indicate the spin orientations of the local magnetic moments. (b, d, f, h) Magnon dispersion curves along the high-symmetry path $\Gamma$-$\text{K}$-$\text{M}$-$\text{K}'$-$\Gamma$ in the first Brillouin zone corresponding to the four aforementioned systems, respectively. The red and blue solid lines denote the spin-up ($\uparrow$) and spin-down ($\downarrow$) magnon band branches, respectively. The insets within each dispersion plot display the symmetry distribution patterns of magnon spin splitting in momentum space within the first Brillouin zone; the red and blue regions represent positive and negative spin splitting, respectively, distinctly manifesting the $p$-wave or $f$-wave splitting features corresponding to the respective systems. The light-field coupling parameters utilized in the calculations are (b) $\lambda=0.2$, (d) $\lambda=0.2$, (f) $\lambda=0.4$, and (h) $\lambda=0.4,\hbar\Omega=1meV$.}
		\label{pic5}
	\end{figure}

	In the main text, we adopt monolayer $\text{MnPS}_3$ as the primary candidate material for realizing odd-parity magnons. As a two-dimensional van der Waals antiferromagnetic material that $\text{MnPS}_3$ has been widely synthesized experimentally~\cite{leeTunnelingTransportMono2016a}, with space group $P\bar{3}1m$ (No. 162). As illustrated in Fig.~\ref{pic5}(a), in the monolayer structure, Mn atoms are arranged in a honeycomb lattice, while P-P dimers reside at the center of the honeycomb, encapsulated above and below by S atom octahedra. Experiments have demonstrated that monolayer $\text{MnPS}_3$ exhibits a Néel-type antiferromagnetic order~\cite{longPersistenceMagnetismAtomically2020}. The local magnetic moment orientations of the Mn atoms are shown in Fig.~\ref{pic5}(a). In our first-principles calculations, we set the effective Coulomb repulsion energy for Mn atoms to $U_{\text{eff}} = 4.0$ eV. Utilizing the aforementioned \texttt{TB2J} scheme, we extracted the exchange coupling constants for the collinear antiferromagnetic ground state. The nearest-neighbor coupling strength is approximately $J_1 \approx 3.5$ meV, and the next-nearest-neighbor coupling strength is approximately $J_2 \approx 0.2$ meV.
	
	Recent experimental studies indicate that macroscopic alloyed $\text{CrVI}_6$ crystals can be synthesized via the chemical vapor transport (CVT) method~\cite{panGrowthHighqualityCrI32022a}. Constrained by current experimental preparation techniques, the V atoms in macroscopic samples are typically randomly distributed, manifesting bulk ferromagnetism. To further substantiate the universality of the bilayer A-type antiferromagnetic model proposed in the main text, we abstract an idealized, periodically ordered AA stacking bilayer model here and assign its interlayer coupling as A-type antiferromagnetic. In this model, the space group symmetry of the non-magnetic lattice diminishes to $P\bar{3}$ (No. 147). As illustrated in Fig.~\ref{pic5}(c), the original $\text{CrI}_3$ unit cell contains two Cr atoms per layer. In our model, the site-1 Cr atom in the bottom-layer unit cell is substituted by a V atom, while in the top layer, the site-1 Cr atom is preserved and the site-2 Cr atom is substituted by a V atom. Because $\text{Cr}^{3+}$ and $\text{V}^{3+}$ possess different outermost electron configurations, they have unequal local magnetic moments, approximately 3 $\mu_B$ and 2 $\mu_B$, respectively. This configuration of unequal magnetic moment magnitudes breaks the $[\mathcal{C}_{2\perp}||\mathcal{M}_z]$ symmetry, yet still preserves the $[\mathcal{C}_{2\perp}||\mathcal{P}]$ symmetry. In our calculations, the $U_{\text{eff}}$ for both Cr and V atoms is set to 3.0 eV.
	
	In the main text, we discussed how to satisfy the symmetry prerequisites for odd-parity magnons by engineering unequal sublattice magnetic moments like the aforementioned $\text{CrVI}_6$. In real two-dimensional van der Waals magnetic materials, apart from relying on unequal magnetic moments, interlayer sliding can analogously break $[\mathcal{C}_{2\perp}||\mathcal{M}_z]$. Bilayer $\text{FeBr}_3$ and bilayer $\text{CrI}_3$ serve as promising candidate experimental platforms in this regard. Bilayer $\text{FeBr}_3$ is a van der Waals magnetic material with an intrinsic A-type antiferromagnetic ground state. Recent experiments have successfully synthesized high-quality bulk $\text{FeBr}_3$ single crystals~\cite{coleExtremeSensitivityMagnetic2023}. Experimental evidence shows that bulk $\text{FeBr}_3$ belongs to the rhombohedral crystal system (space group $R\bar{3}$, No. 148), displays ABC stacking along the $c$-axis, and its magnetic moments are aligned along the $c$-axis, featuring antiparallel interlayer coupling. Given its archetypal layered van der Waals characteristics, it holds the potential to be mechanically exfoliated to the bilayer limit both theoretically and experimentally. We truncate adjacent layers derived from the bulk structure to formulate the bilayer $\text{FeBr}_3$ model. As shown in Fig.~\ref{pic5}(e), the bilayer obtained by truncating the ABC-stacked bulk structure naturally adopts an AB stacking configuration. In this limit, the crystal space group of the bilayer system lowers from the bulk $R\bar{3}$ to $P\bar{3}$ (No. 147). In our calculations, we preserved its bulk-intrinsic A-type antiferromagnetic order and fixed the van der Waals interlayer spacing to the experimentally measured bulk value with the Fe atom $U_{\text{eff}}$ set to 4.0 eV. The stacking sliding, synergized with the antiferromagnetic order, breaks the $[\mathcal{C}_{2\perp}||\mathcal{M}_z]$ symmetry of the system, but preserves inversion symmetry $[\mathcal{C}_{2\perp}||\mathcal{P}]$. Furthermore, since the system retains the $\mathcal{C}_{3z}$ rotational symmetry, its magnons exhibit the $f$-wave spin splitting pattern shown in Fig.~\ref{pic5}(d).
	
	Bulk $\text{CrI}_3$ generally exhibits rhombohedral stacking with interlayer ferromagnetic coupling at low temperatures. For the mechanically exfoliated 2D bilayer $\text{CrI}_3$, experiments have affirmed that its magnetic ground state transitions to A-type antiferromagnetism~\cite{huangLayerdependentFerromagnetismVan2017}. Experimental data reveal that bilayer $\text{CrI}_3$ possesses $C_{2h}$ point group symmetry corresponding to space group $C2/m$, No. 12, verifying that it exhibits interlayer sliding and adopts monoclinic stacking~\cite{sunGiantNonreciprocalSecondharmonic2019}. We constructed a bilayer A-AFM model embodying this monoclinic sliding configuration in our calculations with the Cr atom $U_{\text{eff}} = 3.0$ eV. Different from bilayer $\text{FeBr}_3$ which maintains $C_{3z}$ symmetry, bilayer monoclinic $\text{CrI}_3$ has $[\mathcal{C}_{2\perp}||C_{2[1\bar{1}0]}]$ symmetry. Upon breaking the effective time-reversal symmetry, this system achieves magnon spin splitting with a $p$-wave. Moreover, considering the symmetry of the light field, $[\mathcal{C}_{2\perp}||\mathcal{T}\mathcal{C}_{2[1\bar{1}0]}]$ strictly enforces the degeneracy of the magnon bands along the $\Gamma$-K' path in momentum space. The symmetry constraints are consistent with our calculation results, as shown in Fig.~\ref{pic5}(h).
	
	\bibliography{oddmagnon}